\PassOptionsToPackage{unicode}{hyperref}
\PassOptionsToPackage{naturalnames}{hyperref}

\documentclass[a4paper,11pt]{article}

\usepackage{amsfonts,amssymb,amsmath,amsthm}	
\usepackage[english]{babel}                 	
\usepackage[T1]{fontenc}                    	
\usepackage{setspace}      						
\usepackage{graphicx}      						
\usepackage{tikz}          						
\usepackage{enumitem}
\usepackage[parfill]{parskip}                   
\usepackage[permil]{overpic}       				
\usepackage{fancybox}							
\usepackage{ifthen}								
\usepackage[labelfont=normalfont]{subcaption}
\usepackage{adjustbox}
\usepackage[ plainpages = false, pdfpagelabels,
	     bookmarks,
	     bookmarksopen = true,
	     bookmarksnumbered = true,
	     breaklinks = true,
	     linktocpage,
	     pagebackref,         
	     colorlinks = false,  
         hypertexnames=false,   
	     linkcolor = blue,
	     urlcolor  = blue,
	     citecolor = red,
	     anchorcolor = green,
	     hyperindex = true,
	     hyperfigures
	     ]{hyperref}
\usepackage{tikz}
\usepackage{tikz-3dplot}
\usepackage{fancyhdr, graphicx}
\usepackage{pgfgantt}
\usepackage{bm}

\usepackage[legalpaper, margin=1in]{geometry}
\usepackage{bbm}
\usepackage{lineno}
\usepackage[colorinlistoftodos,prependcaption]{todonotes}
\usepackage{cleveref}
\usepackage{amssymb}
\usepackage{titlesec}
\usepackage{booktabs}
\usepackage{lineno}
\usepackage{siunitx,longtable}
\graphicspath{{figures/}}
\usepackage{pgfplots}
\pgfplotsset{compat=newest}
\usetikzlibrary{plotmarks}
\usetikzlibrary{arrows.meta}
\usepgfplotslibrary{patchplots}
\usepackage{grffile}
\usepackage{amsmath}
\pgfplotsset{plot coordinates/math parser=false}
\newlength\figureheight
\newlength\figurewidth
\pgfplotsset{every axis/.append style={
                    label style={font=\small},
                    tick label style={font=\footnotesize},
                    legend style={font=\tiny},
                    title style={font=\small}
                    }}
\usetikzlibrary{plotmarks}
\pgfplotsset{minor grid style={dotted,gray}} 
\pgfplotsset{every axis/.append style={thick, tick style=semithick}}
\usepgfplotslibrary{groupplots}
\pgfplotsset{xticklabel style={/pgf/number format/fixed,
        /pgf/number format/precision=5},yticklabel style={/pgf/number format/fixed,
        /pgf/number format/precision=5}}

\newcommand{\vect}[1]{\textrm{\boldmath${#1}$}} 
\newcommand{\rare}{\mathbf{R}} 
\newcommand{\hy}{\mathbf{H}} 

\DeclareMathOperator{\diag}{diag}               
\newcommand{\matr}[1]{\mathbf{#1}}              
\newcommand{\norm}[1]{\Vert #1 \Vert}

\newcommand{\Nt}{N_{t}}
\newcommand{\Nc}{N_{c}}
\newcommand{\Nx}{N_{x}}
\newcommand{\Kn}{\mathrm{Kn}}
\newcommand{\email}[1]{\href{mailto:#1}{\texttt{#1}}}

\newcommand{\define}[1]{\textit{#1}}
{\theoremstyle{remark} }

\DeclareMathOperator*{\frepar}{\theta} 
\DeclareMathOperator*{\idhy}{\textbf{\textsf{h}}}
\DeclareMathOperator*{\idrare}{\textbf{\textsf{r}}}
\newcommand{\enc}{g_\mathrm{enc}}
\newcommand{\dec}{g_\mathrm{dec}}
\newcommand{\intrPOD}{p^\mathrm{POD}}
\newcommand{\intrDim}{p^*}
\definecolor{color0}{rgb}{0.12156862745098,0.466666666666667,0.705882352941177}

\binoppenalty=\maxdimen 
\relpenalty=\maxdimen   

\title{Model Order Reduction for the 1D Boltzmann-BGK Equation: Identifying Intrinsic Variables Using Neural Networks}
\author{
Julian Koellermeier\footnote{Bernoulli Institute, University of Groningen \email{j.koellermeier@rug.nl}, corr. author},
Philipp Krah\footnote{Institut de Mathématiques de Marseille, Aix-Marseille Université, \email{philipp.krah@univ-amu.fr}, corr. author},
Julius Reiss\footnote{Technical University of Berlin,
Institute of Fluid Mechanics and Engineering Acoustics},
Zachary Schellin\footnote{Technical University of Berlin,
Institute of Fluid Mechanics and Engineering Acoustics}
}
\begin{document}

\maketitle

\begin{abstract}
Kinetic equations are crucial for modeling non-equilibrium phenomena, but their computational complexity is a challenge. This paper presents a data-driven approach using reduced order models (ROM) to efficiently model non-equilibrium flows in kinetic equations by comparing two ROM approaches: Proper Orthogonal Decomposition (POD) and autoencoder neural networks (AE). While AE initially demonstrate higher accuracy, POD's precision improves as more modes are considered. Notably, our work recognizes that the classical POD-MOR approach, although capable of accurately representing the non-linear solution manifold of the kinetic equation, may not provide a parsimonious model of the data due to the inherently non-linear nature of the data manifold. We demonstrate how AEs are used in finding the intrinsic dimension of a system and to allow correlating the intrinsic quantities with macroscopic quantities that have a physical interpretation.

\end{abstract}

{\bf Keywords}: Model order reduction, data-driven methods, kinetic equations, neural autoencoder networks, proper orthogonal decomposition, Sod shock tube, Boltzmann-BGK

\section{Introduction}
Kinetic equations are widely used in science and engineering \cite{Koellermeier2018b, Maes2023, McClarren2010, Struchtrup2008}. They allow the modeling of deviations from an equilibrium model which is given by an underlying macroscopic equation like the Euler equations, providing detailed insight into fundamental physical processes \cite{Torrilhon2016}. However, kinetic equations are often characterized by a large dimensional phase space, making them computationally expensive to solve and sometimes even unfeasible for realistic applications \cite{Torrilhon2016}.

Investing in solving kinetic equations is only beneficial if large deviations from equilibrium are present \cite{Torrilhon2016}. Striking a balance between a fast but inaccurate equilibrium solver and a slow but accurate non-equilibrium solver remains an open challenge. Our work aims to address this challenge by providing a proof-of-concept for a data-driven solution to efficient modeling of flows in different non-equilibrium regimes.

In the field of non-equilibrium gas flows, several standard methods to discretize the high dimensional phase space exist. Particle-based Monte Carlo methods are only tractable in the free flight regime and out of scope in the transition regime of moderate non-equilibrium unless special techniques are used \cite{Debrabrant2017}. The straightforward Discrete Velocity Method (DVM) uses a pointwise discretization of the velocity space, potentially leading to a large number of equations \cite{Brull2020}. Specially tailored moment models are based on the expansion of the particle distribution function and lead to a set of extended fluid dynamical equations \cite{Torrilhon2016}. However, it is by no means clear a-priori how many equations are sufficient and which variables are optimal \cite{Koellermeier2017a, Torrilhon2015}.

To tackle the computational complexity of kinetic equations, recently, reduced order models (ROM) have been introduced, enabling reductions in computational complexity by orders of magnitudes \cite{bernard2018reduced,Einkemmer2019,Einkemmer2021,EinkemmerHuYing2021}. Two different approaches have been followed in the literature. The classical offline-online decomposition as used by \cite{bernard2018reduced} involves a two-stage procedure. In the offline stage, the full order model (FOM) is assessed to create a database, which is then utilized to generate a data-dependent basis through proper orthogonal decomposition (POD). This basis allows an efficient description of the FOM on a low-dimensional linear subspace during the online phase. On the other hand, the online adaptive basis method called dynamic low-rank approximation \cite{KochLubich2007} constructs the low dimensional linear basis during the online phase itself, eliminating the need to evaluate the expensive FOM. It has been successfully applied to kinetic equations in the works by Einkemmer et al. \cite{Einkemmer2019,Einkemmer2021,EinkemmerHuYing2021}. However, the additional complexity of updating the basis during the evolution makes it less online efficient than the classical offline-online approach shown in \cite{KoellermeierKrahKush2023} for a shallow water moment model.

In this work, we adopt the same offline strategies as in \cite{bernard2018reduced}. Specifically, we sample data for a classical test case called Sod shock tube using a discrete velocity model as our FOM and compare the compression of the linear reduced subspace created by POD with a non-linear description provided by neural autoencoder networks. Neural networks, based on the universal approximation theorem \cite{pinkus_1999}, allow for the approximation of a wide range of function classes and appear promising in identifying the intrinsic dimension of a system. However, the non-linear relation between macroscopic model equations and the discrete velocity model hinders the determination of these dimensions using linear reduction methods like the POD.

This paper aims to utilize these data-driven model reduction techniques to reduce the number of describing variables and equations and determine how many and which variables are useful in specific test cases. For the non-vanishing Knudsen number, we expect to need more non-equilibrium variables with corresponding balance laws, while in the limit of vanishing Knudsen number, we expect to recover the Euler equations, given by conservation laws for mass, momentum, and energy. To the knowledge of the authors, this is the first paper aiming to bridge the gap between equilibrium and non-equilibrium flows using neural networks in this way.

The long-term objective of this line of work is to enable dynamically adapting the model by varying the number of variables during the online phase, paving the way for more efficient and accurate model adaptive simulations of kinetic equations.

The organization of the paper is as follows: In Section 2, we introduce the 1D model problem and the reference data used for model reduction. Section 3 describes the two model reduction techniques used in this study: Proper Orthogonal Decomposition (POD) and Autoencoder Networks. The results are presented in Section 4, and the paper concludes with a summary in Section 5.

\section{The Boltzmann-BGK Model and Data}
\label{sec:BGK}
This paper considers a proof-of-concept of using reduced models for the solution approximation of the 1D Boltzmann-BGK equation \cite{Bhatnagar1954} for monoatomic, ideal gases 
\begin{equation}
    \label{Eq:BGK}
    \partial_t f + c \partial_x f = \frac{1}{\tau} (f_M - f),
\end{equation}
which is a potentially high-dimensional equation for the unknown probability density function $f(t,x,c)$, where $t \in \mathbb{R}^+$ is the time, $x \in \mathbb{R}$ is the spatial variable, and $c \in \mathbb{R}$ the microscopic particle velocity. For simplicity we consider the one-dimensional case in this paper, but the results can be extended to the multi-dimensional case.

Computing solutions and generating data of the Boltzmann-BGK model is essential for industrial and scientific applications, but often so computationally prohibitive that a large number of test cases is not feasible. To reduce time and cost during the data generating process, experiments or numerical simulations can be replaced by reduced-order models (ROMs). 

For standard continuum flows the widely-used Euler equations can be applied, but more rarefied regimes require different extended fluid dynamical models. Rarefaction levels are distinguished with the help of the Knudsen number $\Kn$ defined by the ratio of the mean free path length of the particles $\lambda$ over a reference length $l$ as
\begin{equation}
	\Kn = \frac{\lambda}{l}.
\end{equation}

The right-hand side of the BGK collision operator \eqref{Eq:BGK} models the relaxation with relaxation time $\tau \in \mathbb{R}^+$ towards the equilibrium Maxwellian distribution $f_M(t,x,c)$ given by
\begin{equation}
    f_M(t,x,c) = \frac{\rho(t,x)}{(2\pi R T(t,x))^{\frac{3}{2}}}\exp \left(-\frac{(c - u(t,x))^2}{2 R T(t,x)}\right),
\end{equation}
where $\rho(t,x)$, $v(t,x)$ and $T(t,x)$ are density, bulk velocity, and temperature of the flow, respectively. $R$ is the universal gas constant.
In this work, we consider the relaxation time $\tau$ a parameter and set it equal to the Knudsen number, $\tau = \Kn$, however, the relaxation time can also be changed, e.g., to depend on the gas density and temperature in addition. 

For practical computations, we consider macroscopic moments of the distribution function, which are given by multiplying the distribution function with the co-called collision invariants $(1,c,\frac{1}{2} c^2)$ and integrating in velocity space
\begin{eqnarray}
	\rho(t,x) &=& \int f(t,x,c) \,\mathrm{d}c,\label{Eq:Moments1}\\
	\quad\rho(t,x) u(t,x) &=& \int c f(t,x,c) \,\mathrm{d}c,\label{Eq:Moments2}\\
	\quad E(t,x) &=& \int \frac{1}{2}c^2 f(t,x,c)  \,\mathrm{d}c,\label{Eq:Moments3}
\end{eqnarray}
where $E$ denotes the total energy. The temperature $T(t,x)$ and the pressure $p(t,x)$ can be obtained by
\begin{equation}
	T(t,x) = \frac{2E(t,x)}{3\rho(t,x)} - \frac{u(t,x)^2}{3} \quad\textrm{and}\quad p(t,x) = \rho(t,x) T(t,x).
\end{equation}

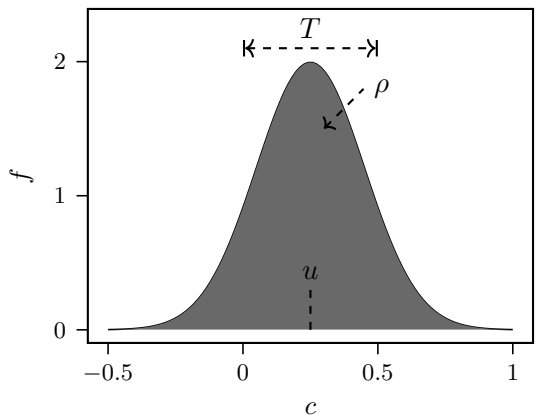
\begin{figure}[htp!]
	\centering
\trimbox{1cm 1.1cm 1cm 0.2cm}{\begin{tikzpicture}

\begin{axis}[
legend cell align={left},
legend style={fill opacity=0.8, draw opacity=1, text opacity=1, draw=white!80!black},
tick align=outside,
tick pos=left,
x grid style={white!69.0196078431373!black},
xlabel={\(c\)},
xmin=-0.575, xmax=1.075,
xtick style={color=black},
y grid style={white!69.0196078431373!black},
ylabel={\(f\)},
ymin=-0.0978129180007683, ymax=2.39285680307655,
ytick style={color=black},
width=0.5\textwidth,
height=0.4\textwidth
]

\addplot [semithick, black]
table {%
	-0.5 0.00176297841183723
	-0.484848484848485 0.00233549990938288
	-0.46969696969697 0.00307623995021814
	-0.454545454545455 0.0040287289903483
	-0.439393939393939 0.00524594088238899
	-0.424242424242424 0.00679182087082491
	-0.409090909090909 0.00874292085377632
	-0.393939393939394 0.0111901101946249
	-0.378787878787879 0.0142403174963827
	-0.363636363636364 0.018018244420721
	-0.348484848484849 0.0226679772544878
	-0.333333333333333 0.0283544060972295
	-0.318181818181818 0.0352643460570684
	-0.303030303030303 0.0436072406856785
	-0.287878787878788 0.0536153162176469
	-0.272727272727273 0.0655430473059348
	-0.257575757575758 0.0796657922473926
	-0.242424242424242 0.0962774595613305
	-0.227272727272727 0.115687079522804
	-0.212121212121212 0.138214174964358
	-0.196969696969697 0.164182856129127
	-0.181818181818182 0.193914604924168
	-0.166666666666667 0.227719764364528
	-0.151515151515151 0.265887808434693
	-0.136363636363636 0.308676534413689
	-0.121212121212121 0.356300391543538
	-0.106060606060606 0.408918233661475
	-0.0909090909090909 0.466620855301653
	-0.0757575757575757 0.529418736525693
	-0.0606060606060606 0.597230476778283
	-0.0454545454545454 0.669872437749538
	-0.0303030303030303 0.747050135179608
	-0.0151515151515151 0.828351915965591
	0 0.91324542694511
	0.0151515151515151 1.00107732368077
	0.0303030303030303 1.09107658129022
	0.0454545454545454 1.18236165636822
	0.0606060606060607 1.2739516126026
	0.0757575757575758 1.36478116780159
	0.0909090909090909 1.45371945330681
	0.106060606060606 1.53959210603912
	0.121212121212121 1.62120614747867
	0.136363636363636 1.69737695187443
	0.151515151515151 1.76695647690227
	0.166666666666667 1.82886183207001
	0.181818181818182 1.88210320029322
	0.196969696969697 1.92581011126961
	0.212121212121212 1.95925509432898
	0.227272727272727 1.98187381354832
	0.242424242424242 1.99328090666395
	0.257575757575758 1.99328090666395
	0.272727272727273 1.98187381354832
	0.287878787878788 1.95925509432898
	0.303030303030303 1.92581011126961
	0.318181818181818 1.88210320029322
	0.333333333333333 1.82886183207001
	0.348484848484849 1.76695647690227
	0.363636363636364 1.69737695187443
	0.378787878787879 1.62120614747867
	0.393939393939394 1.53959210603912
	0.409090909090909 1.45371945330681
	0.424242424242424 1.36478116780159
	0.439393939393939 1.2739516126026
	0.454545454545455 1.18236165636822
	0.46969696969697 1.09107658129022
	0.484848484848485 1.00107732368077
	0.5 0.91324542694511
	0.515151515151515 0.828351915965591
	0.53030303030303 0.747050135179608
	0.545454545454545 0.669872437749538
	0.560606060606061 0.597230476778283
	0.575757575757576 0.529418736525694
	0.590909090909091 0.466620855301653
	0.606060606060606 0.408918233661474
	0.621212121212121 0.356300391543538
	0.636363636363636 0.308676534413689
	0.651515151515152 0.265887808434693
	0.666666666666667 0.227719764364528
	0.681818181818182 0.193914604924168
	0.696969696969697 0.164182856129127
	0.712121212121212 0.138214174964358
	0.727272727272727 0.115687079522804
	0.742424242424242 0.0962774595613305
	0.757575757575758 0.0796657922473926
	0.772727272727273 0.0655430473059348
	0.787878787878788 0.0536153162176469
	0.803030303030303 0.0436072406856785
	0.818181818181818 0.0352643460570684
	0.833333333333333 0.0283544060972294
	0.848484848484849 0.0226679772544878
	0.863636363636364 0.018018244420721
	0.878787878787879 0.0142403174963826
	0.893939393939394 0.0111901101946249
	0.909090909090909 0.00874292085377631
	0.924242424242424 0.00679182087082491
	0.939393939393939 0.00524594088238899
	0.954545454545455 0.0040287289903483
	0.96969696969697 0.00307623995021814
	0.984848484848485 0.00233549990938288
	1 0.00176297841183723
};

\path [draw=none, fill=white!41.1764705882353!black]
(axis cs:-0.5,0.00176297841183723)
--(axis cs:-0.484848484848485,0.00233549990938288)
--(axis cs:-0.46969696969697,0.00307623995021814)
--(axis cs:-0.454545454545455,0.0040287289903483)
--(axis cs:-0.439393939393939,0.00524594088238899)
--(axis cs:-0.424242424242424,0.00679182087082491)
--(axis cs:-0.409090909090909,0.00874292085377632)
--(axis cs:-0.393939393939394,0.0111901101946249)
--(axis cs:-0.378787878787879,0.0142403174963827)
--(axis cs:-0.363636363636364,0.018018244420721)
--(axis cs:-0.348484848484849,0.0226679772544878)
--(axis cs:-0.333333333333333,0.0283544060972295)
--(axis cs:-0.318181818181818,0.0352643460570684)
--(axis cs:-0.303030303030303,0.0436072406856785)
--(axis cs:-0.287878787878788,0.0536153162176469)
--(axis cs:-0.272727272727273,0.0655430473059348)
--(axis cs:-0.257575757575758,0.0796657922473926)
--(axis cs:-0.242424242424242,0.0962774595613305)
--(axis cs:-0.227272727272727,0.115687079522804)
--(axis cs:-0.212121212121212,0.138214174964358)
--(axis cs:-0.196969696969697,0.164182856129127)
--(axis cs:-0.181818181818182,0.193914604924168)
--(axis cs:-0.166666666666667,0.227719764364528)
--(axis cs:-0.151515151515151,0.265887808434693)
--(axis cs:-0.136363636363636,0.308676534413689)
--(axis cs:-0.121212121212121,0.356300391543538)
--(axis cs:-0.106060606060606,0.408918233661475)
--(axis cs:-0.0909090909090909,0.466620855301653)
--(axis cs:-0.0757575757575757,0.529418736525693)
--(axis cs:-0.0606060606060606,0.597230476778283)
--(axis cs:-0.0454545454545454,0.669872437749538)
--(axis cs:-0.0303030303030303,0.747050135179608)
--(axis cs:-0.0151515151515151,0.828351915965591)
--(axis cs:0,0.91324542694511)
--(axis cs:0.0151515151515151,1.00107732368077)
--(axis cs:0.0303030303030303,1.09107658129022)
--(axis cs:0.0454545454545454,1.18236165636822)
--(axis cs:0.0606060606060607,1.2739516126026)
--(axis cs:0.0757575757575758,1.36478116780159)
--(axis cs:0.0909090909090909,1.45371945330681)
--(axis cs:0.106060606060606,1.53959210603912)
--(axis cs:0.121212121212121,1.62120614747867)
--(axis cs:0.136363636363636,1.69737695187443)
--(axis cs:0.151515151515151,1.76695647690227)
--(axis cs:0.166666666666667,1.82886183207001)
--(axis cs:0.181818181818182,1.88210320029322)
--(axis cs:0.196969696969697,1.92581011126961)
--(axis cs:0.212121212121212,1.95925509432898)
--(axis cs:0.227272727272727,1.98187381354832)
--(axis cs:0.242424242424242,1.99328090666395)
--(axis cs:0.257575757575758,1.99328090666395)
--(axis cs:0.272727272727273,1.98187381354832)
--(axis cs:0.287878787878788,1.95925509432898)
--(axis cs:0.303030303030303,1.92581011126961)
--(axis cs:0.318181818181818,1.88210320029322)
--(axis cs:0.333333333333333,1.82886183207001)
--(axis cs:0.348484848484849,1.76695647690227)
--(axis cs:0.363636363636364,1.69737695187443)
--(axis cs:0.378787878787879,1.62120614747867)
--(axis cs:0.393939393939394,1.53959210603912)
--(axis cs:0.409090909090909,1.45371945330681)
--(axis cs:0.424242424242424,1.36478116780159)
--(axis cs:0.439393939393939,1.2739516126026)
--(axis cs:0.454545454545455,1.18236165636822)
--(axis cs:0.46969696969697,1.09107658129022)
--(axis cs:0.484848484848485,1.00107732368077)
--(axis cs:0.5,0.91324542694511)
--(axis cs:0.515151515151515,0.828351915965591)
--(axis cs:0.53030303030303,0.747050135179608)
--(axis cs:0.545454545454545,0.669872437749538)
--(axis cs:0.560606060606061,0.597230476778283)
--(axis cs:0.575757575757576,0.529418736525694)
--(axis cs:0.590909090909091,0.466620855301653)
--(axis cs:0.606060606060606,0.408918233661474)
--(axis cs:0.621212121212121,0.356300391543538)
--(axis cs:0.636363636363636,0.308676534413689)
--(axis cs:0.651515151515152,0.265887808434693)
--(axis cs:0.666666666666667,0.227719764364528)
--(axis cs:0.681818181818182,0.193914604924168)
--(axis cs:0.696969696969697,0.164182856129127)
--(axis cs:0.712121212121212,0.138214174964358)
--(axis cs:0.727272727272727,0.115687079522804)
--(axis cs:0.742424242424242,0.0962774595613305)
--(axis cs:0.757575757575758,0.0796657922473926)
--(axis cs:0.772727272727273,0.0655430473059348)
--(axis cs:0.787878787878788,0.0536153162176469)
--(axis cs:0.803030303030303,0.0436072406856785)
--(axis cs:0.818181818181818,0.0352643460570684)
--(axis cs:0.833333333333333,0.0283544060972294)
--(axis cs:0.848484848484849,0.0226679772544878)
--(axis cs:0.863636363636364,0.018018244420721)
--(axis cs:0.878787878787879,0.0142403174963826)
--(axis cs:0.893939393939394,0.0111901101946249)
--(axis cs:0.909090909090909,0.00874292085377631)
--(axis cs:0.924242424242424,0.00679182087082491)
--(axis cs:0.939393939393939,0.00524594088238899)
--(axis cs:0.954545454545455,0.0040287289903483)
--(axis cs:0.96969696969697,0.00307623995021814)
--(axis cs:0.984848484848485,0.00233549990938288)
--(axis cs:1,0.00176297841183723)
--cycle;
\addlegendimage{area legend, draw=none, fill=white!41.1764705882353!black}
\draw [dashed,|<->|] (axis cs:0,2.1) -- (axis cs:0.5,2.1);
\draw [] (axis cs:0.25,2.1) node[above] {\(T\)};
\draw [dashed] (axis cs:0.25,0)--(axis cs:0.25,0.3);
\draw [] (axis cs:0.25,0.3) node[above] {\(u\)};
\draw [dashed,<-] (axis cs:0.3,1.5)-- +(15pt,15pt) node[right] {\(\rho\)};
\end{axis}
\end{tikzpicture}}
	\caption{Illustration of the macroscopic moments corresponding to an example distribution function. }
	\label{Fig:Demo Macro}
\end{figure}

\Cref{Fig:Demo Macro} illustrates the relation between the macroscopic moments and the distribution function $f(t,x,c)$ at a certain position in time and space. The density $\rho(t,x)$ is the integral of the distribution function, which is centered around the macroscopic velocity $u(t,x)$, and the mean deviation is related to the temperature $T(t,x)$.

The Boltzmann-BGK equation \eqref{Eq:BGK} is in equilibrium when $f = f_M$. Multiplying the equilibrium solution with the collision invariants and integrating in velocity space, one finds the Euler equations of classical gas dynamics
\begin{align}
	\partial_t&\rho + \partial_x(\rho u) = 0, \label{Eq:Conservation1} \\
	\partial_t&(\rho u) + \partial_x(\rho u^2 + p) = 0, \label{Eq:Conservation2}\\
	\partial_t&E + \partial_x(u(E+p)) = 0, \label{Eq:Conservation3}
\end{align}
which are conservation laws for mass, momentum, and energy, respectively.

For distribution functions further away from equilibrium, for example due to a larger relaxation time $\tau$ and a significantly large Knudsen number $\Kn$, the Euler equations do not give accurate results. In this case, additional equations can be used, which are derived by the so-called method of moments \cite{Koellermeier2017d,Torrilhon2016}. This effectively leads to an extended set of equations, called moment model. It is possible to preserve important properties like hyperbolicity with moment models \cite{Fan2016,Koellermeier2014}. The additional equations (for example for the heat flux and higher-order moments) add complexity, but allow for more accurate solutions \cite{Koellermeier2017b,Torrilhon2015}. However, it is often unclear a-priori, how many equations are needed for an efficiently accurate and computationally feasible solution. In this work, we aim to give a proof-of-concept for a data-based identification of the necessary number of variables, called the intrinsic physical dimension.

\subsection{Sod shock tube test case and reference data}
\label{sec:test_and_ref_data}
Sod's shock tube is a well-established test case in the field of rarefied gases \cite{Koellermeier2017a}. It uses discontinuous initial conditions based on equilibrium values
\begin{equation}
    \label{eqn:shock_tube_problem}
    \begin{cases} (\rho_L, u_L, p_L) = (1,0,1) &\mbox{if } x < 0.5, \\ (\rho_R, u_R, p_R) = (0.125,0,0.1) & \mbox{if } x > 0.5, \end{cases}
\end{equation}
corresponding to a jump in density and pressure at $x=0.5$ due to a diaphragm at that position, which is removed at time $t=0$.

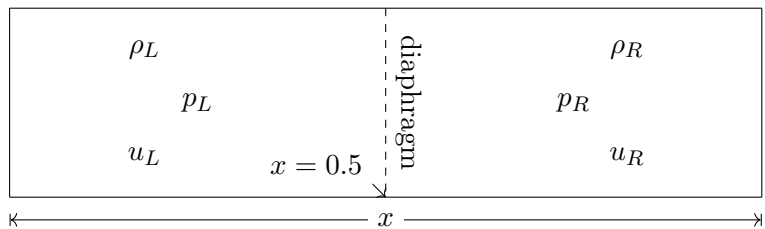
\begin{figure}[htp!]
	\centering
	\begin{tikzpicture}
\node[] at (0, 0) (a) {};
\node[] at (0,-.3) (a1){};
\node[] at (-.3,0) (a2){};
\node[] at (10,0) (b) {};
\node[] at (10,-.3) (b1){};

\node[] at (0,2.5)  (c) {};
\node[] at (-.3,2.5) (c1){};
\node[] at (10,2.5) (d) {};

\node[] at (5,2.5)  (e) {};
\node[] at (5,0) (f) {};

\draw[] (a.center) -- (b.center) {};
\draw[|<->|] (a1.center) -- (b1.center) node[midway,fill=white] (TextNode) {\(x\)};
\draw[] (a.center) -- (c.center) {};
\draw[] (c.center) -- (d.center) node[above] {};
\draw[] (d.center) -- (b.center) node[above] {};
\draw[dashed] (e.center) -- (f.center) node[sloped,above,midway] {diaphragm};
\draw[<-] (f.center)-- +(-5pt,5pt) node[above left] {\(x=0.5\)};

\node[] at (2.5,1.25) (T){\(p_L\)};
\node [above left of=T] (rho) {\(\rho_L\)}; 
\node [below left of=T] (v) {\(u_L\)};

\node[] at (7.5,1.25) (T0){\(p_R\)};
\node [above right of=T0] (rho0) {\(\rho_R\)}; 
\node [below right of=T0] (v0) {\(u_R\)};
\end{tikzpicture}
	\caption{Problem setup for the 1D Sod shock tube. A diaphragm at the center is initially separating the domain in two regions, where initial conditions for density $\rho$, macroscopic velocity $u$, and pressure $p$ are indicated.}
	\label{Fig:SodProbSetup}
\end{figure}

The problem setup at $t=0$ is shown in \cref{Fig:SodProbSetup}, which is split into two regions left and right of the diaphragm. 

For the generation of reference data we employ a discrete velocity method (DVM) \cite{Mieussens2000}, which uses a pointwise microscopic velocity space discretization
\begin{equation}
	\partial_t f_{k}(t,x) = -(c_k) \partial_x f_{k}(t,x) + \frac{1}{\tau}\left({M_f}_{k}(t,x) - f_{k}(t,x)\right),
	\label{Eq:Discrete BGK}
\end{equation}
where a uniform grid in velocity space is considered with $c_k = k\Delta c$ to discretize the distribution function $f_k(t,x) = f(t,x,c_k)$, for some $k \in \mathbb{Z}$. After a subsequent discretization in space, the DVM \cref{Eq:Discrete BGK} leads to a coupled ODE system in time than can be solved with standard methods. 

For the numerical reference data, we use $N_x = 200$ spatial points $x_k \in [0, 1]$, $N_c = 40$ discrete velocities $c_j \in [-10, 10]$ and $N_t = 25$ time steps $t^n \in [0, 0.12]$ summarized in \cref{Tab:Setup}. It is possible to choose another range for the discrete velocity points, but in typical applications the range of the bulk velocity is not known, such that one has to include a safety margin. We therefore chose the domain $[-10, 10]$. The goal of the model order reduction is now to reduce the complexity of the computation using lower dimensional models. For that matter, it is not relevant what the actual error of the numerical reference data is or if is fully converged. It is fair to say that a full reference solution might easily take into account more spatial points, time steps, and discrete velocities, which makes it even more necessary to reduce the complexity. 

For the model order reduction later, we consider two different Knudsen numbers for Sod's shock-tube test case: $\textrm{Kn} = 0.00001$ for a small Knudsen number in the hydrodynamic regime and $\textrm{Kn} = 0.01$ for a relatively large Knudsen number in the rarefied regime.

To understand the behavior of the reference solutions for non-vanishing Knudsen numbers, we first describe the solution for vanishing Knudsen number in equilibrium, i.e., $\Kn = 0$, which can be obtained using the method of characteristics and the Rankine Hugoniot jump conditions connecting the states before and after the shocks \cite{LeVeque2002}.

\begin{figure}[htp!]
	\begin{subfigure}{.46\linewidth}
		\centering
		\begin{tikzpicture}[scale=0.9]
	\node [] at (0,0) (a0) {};
	\node [] at (3.75,0) (a3) {};
	\node [] at (7.5,0) (a6) {};
	\node [on grid,below = 1.5ex of a3] (xd) {$x=0.5$};
	\node [] at (0,1.6) (f0) {};
	\node [] at (3.75,1.6) (f1) {};
	\node [] at (7.5,0.5) (b6) {};
	
	\node [] at (0,2.5) (h0) {};
	\node [] at (3.75,2.5) (h1) {};
	\node [] at (3.75,0.5) (b3) {};
	\node [] at (7.5,1) (d6) {};
	\node [] at (3.75,1) (d3) {};	
	\node [] at (0,3) (i0) {};
	\node [] at (3.75,3) (i3) {};
	\node [] at (7.5,3) (i6) {};
	
	\draw[|-,thick] (a0.center) -- (a3.center) node [midway,above] {$u_1=0$};
	\draw[thick] (f0.center) -- (f1.center) node [midway,above] {$\rho_1$};
	\draw[thick] (h0.center) -- (h1.center) node [midway,above] {$p_1$};
	
	\draw[->,thick] (a3.center) -- (a6.center) node [midway,above] { $u_R=0$} node [below left] {$x$};
	\draw[thick] (b3.center) -- (b6.center) node [midway,above] {$\rho_R$};
	\draw[thick] (d3.center) -- (d6.center) node [midway,above] {$p_R$};
	\draw[dashed] (a3.center) -- (i3.center) {};
	
\end{tikzpicture}
		\caption{Sod's shock tube ($t=0$) containing two regions with corresponding initial conditions.}
		\label{Fig:SodTime}
	\end{subfigure}\hfill
	\begin{subfigure}{.46\linewidth}
		\centering
		\begin{tikzpicture}[scale=0.9]
	\node [] at (0,0) (a0) {};
	\node [] at (2,0) (a1) {};
	\node [] at (3,0) (a2) {};
	\node [] at (3.5,0) (a3) {};
	\node [] at (4.5,0) (a4) {};
	\node [] at (5.5,0) (a5) {};
	\node [] at (7.5,0) (a6) {};
	\node [on grid,below = 1.5ex of a1] (x1) {$x_1$};
	\node [on grid,below = 1.5ex of a3] (x2) {$x_2$};
	\node [on grid,below = 1.5ex of a4] (x3) {$x_3$};
	\node [on grid,below = 1.5ex of a5] (x4) {$x_4$};
	\node [] at (7.5,0.5) (b6) {};
	\node [] at (5.5,0.5) (b5) {};
	\node [] at (3,.9) (c2) {};
	\node [] at (3.5,.9) (c3) {};
	\node [] at (4.5,.9) (c4) {};
	\node [] at (7.5,1) (d6) {};
	\node [] at (5.5,1) (d5) {};
	\node [] at (4.5,1.5) (e4) {};
	\node [] at (5.5,1.5) (e5) {};
	\node [] at (0,1.6) (f0) {};
	\node [] at (2,1.6) (f1) {};
	\node [] at (3,2) (g2) {};
	\node [] at (3.5,2) (g3) {};
	\node [] at (4.5,2) (g4) {};
	\node [] at (5.5,2) (g5) {};
	\node [] at (7.5,2) (g6) {};
	\node [] at (0,2.5) (h0) {};
	\node [] at (2,2.5) (h1) {};
	\node [] at (3,2.5) (h2) {};
	\node [] at (3.5,2.5) (h3) {};
	\node [] at (4.5,2.5) (h4) {};
	\node [] at (5.5,2.5) (h5) {};
	\node [] at (0,3) (i0) {};
	\node [] at (2,3) (i1) {};
	\node [] at (3,3) (i2) {};
	\node [] at (3.5,3) (i3) {};
	\node [] at (4.5,3) (i4) {};
	\node [] at (5.5,3) (i5) {};
	\node [] at (7.5,3) (i6) {};
	\draw [|->,thick] (a0.center) -- (a6.center) node [below left] {$x$};
	\path [] (a0.center) -- (a1.center) node [midway,above] {$u_L$};
	\path [] (a5.center) -- (a6.center) node [midway,above] { $u_R$};
	\path[] (i0.center) -- (i1.center) node [midway,above] {\tiny 1};
	\path[] (i1.center) -- (i3.center) node [midway,above] {\tiny 2};
	\path[] (i3.center) -- (i4.center) node [midway,above] {\tiny 3};
	\path[] (i4.center) -- (i5.center) node [midway,above] {\tiny 4};
	\path[] (i5.center) -- (i6.center) node [midway,above] {\tiny 5};
	\draw[thick] (b5.center) -- (b6.center) {};
	\path[] (b5.center) -- (b6.center) node [midway,above] {$\rho_R$};
	\path[] (c3.center) -- (c4.center) node [midway,above] {$\rho_3$};
	\draw[thick] (d5.center) -- (d6.center) {};
	\path[] (d5.center) -- (d6.center) node [midway,above] {$p_R$};
	\draw[thick] (e4.center) -- (e5.center) {};
	\path[] (e4.center) -- (e5.center) node [midway,above] {$\rho_4$};
	\draw[thick] plot [smooth] coordinates {(f1)(2.6,1.1)(c2)(c3)(c4)};
	\draw[thick] (f0.center) -- (f1.center) node [midway,above] {$\rho_1$};
	\path[] (g3.center) -- (g4.center) node [midway,above] {$p_3$};
	\path[] (g4.center) -- (g5.center) node [midway,above] {$p_4$};
	\draw[thick] plot [smooth] coordinates {(h1)(2.5,2.2)(g2)(g3)(g4)(g5)};
	\draw[thick] (h3.center) -- (a1.center) {};

	\draw[thick] (h0.center) -- (h1.center) node [midway,above] {$p_1$};
	\draw[thick] (h3.center) -- (h4.center) node [midway,above] {$u_3$};
	\draw[thick] (h4.center) -- (h5.center) node [midway,above] {$u_4$};
	\draw[dashed] (a1.center) -- (i1.center) {};
	\draw[dashed] (a3.center) -- (i3.center) {};
	\draw[dashed] (a4.center) -- (i4.center) {};
	\draw[dashed] (a5.center) -- (i5.center) {};

\end{tikzpicture}
		\caption{Sod's shock tube ($t>0$) containing the rarefaction wave, the contact discontinuity and the shock wave.}
		\label{Fig:SodTime0}
	\end{subfigure}\\ \vfill
	\begin{subfigure}{\textwidth}
		\centering
\begin{tikzpicture}

\begin{groupplot}[group style={group size=3 by 2,horizontal sep=1.1cm,vertical sep=1cm}]
\nextgroupplot[
tick align=outside,
tick pos=left,
x grid style={white!69.0196078431373!black},
xmin=.0025,xmax=0.9975,
xtick style={color=black},
y grid style={white!69.0196078431373!black},
ymin=-10,ymax=10,
ytick style={color=black},
width = .34\textwidth,
height = .2\textwidth,
ylabel= $c$,
y label style={yshift=-1.5em},
x label style={yshift=.3em},
xlabel = $x$
]
\addplot graphics [includegraphics cmd=\pgfimage,xmin=.0025,xmax=0.9975, ymin=-10,ymax=10] {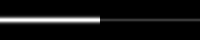};
\node[fill=white] at (axis cs:.83,-6.5) {\small $t=0$};

\nextgroupplot[
tick align=outside,
tick pos=left,
x grid style={white!69.0196078431373!black},
xmin=.0025,xmax=0.9975,
xtick style={color=black},
y grid style={white!69.0196078431373!black},
ymin=-10,ymax=10,
ytick style={color=black},
width = .34\textwidth,
height = .2\textwidth,
ylabel= $c$,
y label style={yshift=-1.5em},
x label style={yshift=.3em},
xlabel = $x$
]
\addplot graphics [includegraphics cmd=\pgfimage,xmin=.0025,xmax=0.9975, ymin=-10,ymax=10] {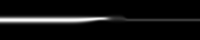};
\node[fill=white] at (axis cs:.775,-6.5) {\small $t=0.06$};

\nextgroupplot[
colorbar,
colorbar style={width=1.5ex,ylabel={}},
colormap/blackwhite,
point meta max=0.398587408411986,
point meta min=1.61655951012117e-88,
tick align=outside,
tick pos=left,
x grid style={white!69.0196078431373!black},
xmin=.0025,xmax=0.9975,
xtick style={color=black},
y grid style={white!69.0196078431373!black},
ymin=-10,ymax=10,
ytick style={color=black},
width = .34\textwidth,
height = .2\textwidth,
ylabel= $c$,
y label style={yshift=-1.5em},
x label style={yshift=.3em},
xlabel = $x$
]
\addplot graphics [includegraphics cmd=\pgfimage,xmin=.0025,xmax=0.9975,ymin=-10,ymax=10] {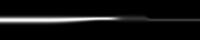};
\node[fill=white] at (axis cs:.775,-6.5) {\small $t=0.12$};

\nextgroupplot[
tick align=outside,
tick pos=left,
x grid style={white!69.0196078431373!black},
xmin=.0025,xmax=0.9975,
xtick style={color=black},
y grid style={white!69.0196078431373!black},
ymin=-10,ymax=10,
ytick style={color=black},
width = .34\textwidth,
height = .2\textwidth,
ylabel= $c$,
y label style={yshift=-1.5em},
x label style={yshift=.3em},
xlabel = $x$
]
\addplot graphics [includegraphics cmd=\pgfimage,xmin=.0025,xmax=0.9975, ymin=-10,ymax=10] {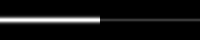};
\node[fill=white] at (axis cs:.83,-6.5) {\small $t=0$};

\nextgroupplot[
tick align=outside,
tick pos=left,
x grid style={white!69.0196078431373!black},
xmin=.0025,xmax=0.9975,
xtick style={color=black},
y grid style={white!69.0196078431373!black},
ymin=-10,ymax=10,
ytick style={color=black},
width = .34\textwidth,
height = .2\textwidth,
ylabel= $c$,
y label style={yshift=-1.5em},
x label style={yshift=.3em},
xlabel = $x$
]
\addplot graphics [includegraphics cmd=\pgfimage,xmin=.0025,xmax=0.9975, ymin=-10,ymax=10] {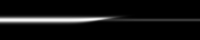};
\node[fill=white] at (axis cs:.775,-6.5) {\small $t=0.06$};
\nextgroupplot[
colorbar,
colorbar style={width=1.5ex,ylabel={}},
colormap/blackwhite,
point meta max=0.406101604565777,
point meta min=1.62163282651898e-88,
tick align=outside,
tick pos=left,
x grid style={white!69.0196078431373!black},
xmin=.0025,xmax=0.9975,
xtick style={color=black},
y grid style={white!69.0196078431373!black},
ymin=-10,ymax=10,
ytick style={color=black},
width = .34\textwidth,
height = .2\textwidth,
ylabel= $c$,
y label style={yshift=-1.5em},
x label style={yshift=.3em},
xlabel = $x$
]
\addplot graphics [includegraphics cmd=\pgfimage,xmin=.0025,xmax=0.9975, ymin=-10,ymax=10] {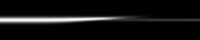};
\node[fill=white] at (axis cs:.775,-6.5) {\small $t=0.12$};
\end{groupplot}

\end{tikzpicture}
		\caption{Discrete reference solutions $f(t,x,c)$ for $\Kn=0.00001$ (top row) and $\Kn=0.01$ (bottom row), for times $t_0=0$ (left), $t_1=0.06$ (center) and $t_2=0.12$ (right).}
		\label{Fig:ExamplesSod}
	\end{subfigure}\\ \vspace{1pt}
 
	\begin{subfigure}{\textwidth}
		\input{MacroFOMhyvsrare.tex}
		\caption{Macroscopic quantities $\rho$, $\rho u$, $E$ ($t=0.12$) for $\Kn=0.00001$ ($\hy$) and $\Kn=0.01$ ($\rare$).}
		\label{Fig:SODHyRare}
	\end{subfigure}\caption{Sod shock tube and reference data. Initial conditions (a); equilibrium solution (b); Reference solutions in rarefied and hydrodynamic regime (c); Macroscopic quantities at $t=0.12s$ (d).}
	\label{Fig:Sod}
\end{figure}

Starting from the initial condition in \cref{Fig:SodTime}, the solution evolves for $t> 0$ and five regions are formed that are depicted in \cref{Fig:SodTime0} \cite{Sod}. A rarefaction wave is moving to the left between $x_1$ and $x_2$. The contact discontinuity is located at $x_3$, where the macroscopic velocity $u$ and the pressure $p$ are continuous in contrast to the density $\rho$ and the energy $E$. $x_4$ is the position of the shock wave.

In non-equilibrium, i.e., for solutions evolving with Knudsen numbers $\Kn > 0$, the solution does not have discontinuities due to the finite relaxation time $\tau$. \Cref{Fig:ExamplesSod} shows the reference solutions $f(t,x,c)$ at $t_0=0$, $t_1=0.06$ and $t_3=0.12$ for the two levels of rarefaction considered in this paper: $\Kn = 0.00001$ and $\Kn =0.01$. Increasing the Knudsen number leads to a smoother transition from region 1 to region 5 with a less pronounced shock front.

\begin{table}[htp]
	\centering
	\caption{Problem setup for the Boltzmann-BGK model in Sod's shock tube.}
	\begin{tabular*}{15cm}{ @{\extracolsep{\fill}} c c c c @{} }
		\toprule
		Variable   & Number of nodes \(i\)&   Domain extension& Step size (uniform)\\   
		\hline
		\(x\) 		&	200&     [0, 1]&	    0.005\\
		\(c\)       &   40 &  		    [-10,10]&	    \(\approx\) 0.51282051\\
		\(t\)   	&	25 &        	[0,0.12]&	      0.005\\
		\bottomrule
	\end{tabular*} \label{Tab:Setup}
\end{table}

\section{Methods}
\label{sec:methods}
In this section we present two common methods used for reducing the dimensionality of high dimensional data: (1) the proper orthogonal decomposition (POD) and (2) neural autoencoder networks (AE). The methods will be used to parameterise the high dimensional data stemming from the DVM simulation, using a linear mapping in case of the POD and a non-linear mapping in case of AE. Although the classical POD-MOR approach shows that linear mappings are sufficient to describe the non-linear solution manifold of the BGK equation to a good accuracy \cite{bernard2018reduced}, it is in general not sufficient to determine a parsimonious model of the full model data, since the data manifold can be non-linear. Here, neural autoencoder networks can be used as they are capable to find the intrinsic dimension of a system. 

\subsection{Proper orthogonal decomposition}

The \define{proper orthogonal decomposition} \cite{SiL1987} (POD) approximates the data with help of dyadic pairs:
\begin{equation}
    \label{eq:POD-approx}
 f(x,t,c_i) \approx \sum_{k=1}^r \hat{f}_k(x,t)\psi_k(c_i) \qquad \text{for } r\ll N_c\,.
\end{equation}
The pairs $\{(\hat{f}_k(x,t),\psi_k(c_i))\}_{k=1,\dots,r}$ are the structures in the data that contain the most energy and they are chosen to minimize the gap between the data and the reconstruction \cref{eq:POD-approx}.
In the following, $\psi_k(c_i)$ are termed POD-modes and $\hat{f}_k(x,t)$ the corresponding reduced variables.

For notation, we define $f^{(i)}(x,t) = f(x,t,c_i)$ and the vectors $f(x,t) = (f^{(1)}(x,t), \dots, f^{(\Nc)}(x,t))$ and $\psi_k = (\psi_k(c_1),\dots, \psi_k(c_{\Nc}))$.
The proper orthogonal decomposition computes the solution of the minimization problem:
\begin{equation}
    \min_{\psi_k}\norm{f(x,t)-\sum_{k=1}^r\langle f(x,t),\psi_k\rangle\psi_k}_2^2\, \quad\text{such that}\quad\langle \psi_k,\psi_l\rangle =\delta_{kl}.
\end{equation}
Technically one can solve this optimization problem using a singular value decomposition (SVD) of the so-called snapshot matrix \cite{KunischVolkwein1999}:
 \begin{equation}
    \label{eq:snapshotmatrix}
                \mathbf{F}=\begin{bmatrix}
                    f^{(1)}(x_1, t_1) &   \cdots & f^{(\Nc)}(x_1, t_1) \\
                    \vdots &  \ddots & \vdots \\
                    f^{(1)}(x_{\Nx}, t_1) &   \cdots & f^{(\Nc)}(x_{\Nx}, t_1) \\
                    f^{(1)}(x_1, t_2) &   \cdots & f^{(\Nc)}(x_1, t_2) \\
                    \vdots & \ddots &  \vdots \\
                    f^{(1)}(x_{\Nx}, t_{\Nt}) &   \cdots & f^{(\Nc)}(x_m, t_{\Nt}) \\
                \end{bmatrix}\in\mathbb{R}^{(\Nx\Nt)\times\Nc}\,.
\end{equation}
The snapshot matrix collects the time and space discrete distribution function in its columns. Each column holds the time-spatial values for a different discrete velocity. Performing an SVD factorizes $\matr{F}$ as
\begin{equation}
    \matr{F}=\matr{\Phi\Sigma \Psi^T},
\end{equation}
with diagonal matrix $\matr{\Sigma}=\diag(\sigma_1,\dots,\sigma_m)$, $m=\min(\Nx\Nt,\Nc)$, containing the singular values $\sigma_1\ge \sigma_2\ge\dots\ge \sigma_m\ge0$ and $\matr{\Phi}\in\mathbb{R}^{\Nx\Nt\times m}, \matr{\Psi}\in\mathbb{R}^{\Nc\times m}$ are orthogonal matrices containing the left and right singular vectors, respectively. 

The first $r$ columns of the truncated $\matr{\Psi}_r=[\psi_1,\dots,\psi_r]\in\mathbb{R}^{\Nc\times r}$ contain the POD modes in \cref{eq:POD-approx}. Together with $\matr{\Sigma}_r=\diag(\sigma_1,\dots,\sigma_r)$ and the $r$ leading left singular vectors $\matr{\Phi}_r\in\mathbb{R}^{(\Nx\Nt)\times r}$ they yield the rank $r$-term approximation $\matr{F}_r$ of the snapshot matrix given by
\begin{equation}
    \matr{F}_r:= \matr{\Phi}_r\matr{\Sigma}_r \matr{\Psi}_r^T.
\end{equation}
According to the Eckart-Young-Mirsky theorem \cite{EcY1936,Mirsky1960} $\matr{F}_r$ is the best rank $r$ approximation and the resulting error in the Frobenius norm is rigorously computed from the trailing singular values
\begin{equation}
    \norm{\matr{F}-\matr{F}_r}_\mathrm{F}^2=\sum_{k=r+1}^m\sigma_k^2\,.
\end{equation}
A common choice for $r$ is to truncate after a certain energy percentage is reached in the reduced system compared to the full system:
\begin{equation}\label{e:ePOD}
    E_\mathrm{cum}=\frac{\norm{\matr{F}_r}_\mathrm{F}}{\norm{\matr{F}}_\mathrm{F}}=\frac{\sum_{k=1}^r\sigma_k^2}{\sum_{k=1}^m\sigma_k^2}.
\end{equation}

In this paper, POD is used to compare with the autoencoder from the next section.

\subsection{Autoencoders}
Neural networks, particularly autoencoder networks, have become widely used tools for dimension reduction \cite{VanDerMaatenPostmaHerik2009}. A comprehensive introduction to autoencoder networks can be found in \cite{Goodfellow-et-al-2016}. Here we only summarize briefly the common idea of autoencoder networks and give the specific details of our implementation.

An autoencoder aims to reproduce the input data while compressing it through an information bottleneck. It consists of two main components:
\begin{itemize}
    \item 
    The encoder, denoted as $\enc$, maps the input data ${f}$ to points ${\hat{f}}$ in a lower-dimensional latent space: $\enc \colon \mathbb{R}^{M}\to\mathbb{R}^r, {f}\mapsto {\hat{f}} = \enc({f})$, $r\ll M$.
    \item 
    The decoder, denoted as $\dec$, reconstructs the input space from the latent representation: $\dec \colon\mathbb{R}^{r}\to\mathbb{R}^{M}, {\hat{f}} \mapsto \dec({\hat{f}}) = {\tilde{f}}$, $r\ll M$.    
\end{itemize}
Note that the dimension of the latent space is denoted by $r$, to match the rank of the POD approximation. 

The autoencoder is defined as the composition of both parts: ${\tilde{f}} = \dec \left( \enc \left( {f} \right) \right)$. 
For our purpose, we identify the discrete velocity space as the input dimension $M=\Nc$. Thus, the autoencoder maps each time-spatial value of the distribution function $f(x,t)\in \mathbb{R}^{\Nc}$ onto a smaller latent space $\hat f(x,t)\in \mathbb{R}^{r}$, which parameterizes the necessary physical information of the system.

The goal of the optimization procedure is to determine $\dec$ and $\enc$ such that the reconstruction error over a set of training/testing data contained in $\matr{F}$ is minimized. The reconstruction error is defined as:
\begin{equation}
    \label{eq:loss}
    \mathcal{L} = \frac{1}{\Nc}\norm{f(x,t)- \dec\circ\enc\circ f(x,t)}_2^2   \,. 
\end{equation}
The reconstruction error is the sum of the two-norm of the discrete velocities vector of the difference between the input data $f$ and the reconstructed data $\tilde{f}$ that has been squeezed through the informational bottleneck. The assumption is, that if the original data can be represented well while the information went through a smaller latent space, there exists a physical law in the latent space that describes the system sufficiently. The intrinsic latent dimension $r=\intrDim$ which is sufficient to describe the data is then called the \define{intrinsic physical dimension} similar to the intrinsic dimension defined in \cite{LeeCarlberg2020}. Such a reduced model is then termed parsimonious because it explains the data with a minimum number of variables.

In the training procedure, the functions $\enc$ and $\dec$ are determined by trainable parameters of the network, referred to as weights and biases. The networks are constructed using a composition of layers $\enc=L_1\circ L_2 \circ \dots \circ L_N$. Typically, each layer $L_n\colon \mathbb{R}^i \to \mathbb{R}^o$ in the network consists of an affine linear mapping $\vect{x}\mapsto h_n(\matr{W}_n {x}+{b}_n)$, where $\matr{W}_n\in\mathbb{R}^{o,i}$ represent the weights, $\vect{b}_n \in\mathbb{R}^o$ denote the biases, and $h_n$ are predefined non-linear functions. The configuration of the input and output dimensions $i$ and $o$ for each layer, the choice of activation function, and the number of layers collectively determine the architecture of the network. The choice of these so-called hyper-parameters is often difficult and a matter of trial and error. 

\paragraph{Architecture}
In our studies we have exploited \define{fully connected neural autoencoder networks} (FCNN) and convolutional autoencoder networks (CNN). However, in this manuscript we restrict ourselves to the results of the fully connected network, since it gave structurally the best results. 
We have studied a variety of different activation functions, hidden layers, batch sizes and depths of the network. The best results concerning the validation error and acceptable training time where obtained by the network
defined in \cref{tab:autoencoder-params}. A comprehensive study of the parameter optimization is attached to the manuscript in \cref{appendix}.

\begin{table}[htp!]
\centering
\begin{tabular}{lcc}
\toprule
\textbf{Parameter} & {hydrodynamic} & {rarefied} \\
\midrule
Layer Sizes & [$\Nc$, 30, $r$, 30, $\Nc$]& [$\Nc$, 40, $r$, 40, $\Nc$] \\
Activation function & ELU & ReLU \\
Loss function & MSE \cref{eq:loss} & MSE \cref{eq:loss}\\
Optimizer & Adam & Adam\\
Learning rate & $10^{-5}$ & $10^{-5}$ \\
Epochs & 3000 & 3000 \\
Batch size & 4 & 4 \\
\bottomrule
\end{tabular}
\caption{Hyper-parameters of Fully Connected Autoencoder Network (FCNN) for the hydrodynamic and rarefied regime.}
\label{tab:autoencoder-params}
\end{table}

\paragraph{Training}
Before the training we initialize the weights of the network using the standard initialization implemented in \texttt{pytorch}. Thus the weights are randomly uniform distributed between $m^{-1/2}$ and $m^{1/2}$ with $m$ being the number of input nodes in the layer.
Our network is trained by splitting the data consisting of $\Nx\times\Nt$ samples in a testing and training set with a 80/20 split over 3000 epochs using a batch size of 4. In each epoch the network is updated using the Adam optimizer with a learning rate of $10^{-5}$. More information about hyperparameters and training of the network can be found in the appendix \cref{appendix}.
 
\section{Results}
\label{sec:results}
We reconstruct the full order model (FOM) solution with the help of POD and an autoencoder, the FCNN, for which the selection of hyperparameters and the training are described in the previous section. Note that we apply both model reduction techniques to reconstruct both the rarefied reference data and the hydrodynamic reference data. The goal is to later determine the intrinsic dimension of the data for both cases.
We therefore compare the two dimension reduction techniques by means of different measures. The intrinsic variables obtained from POD and the FCNN will be referred to as \(\idhy\) and \(\idrare\), where the former describes the intrinsic variables when reducing the hydrodynamic data and the latter when reducing the rarefied data.

For the purpose of comparing the results we define the $L_2$-error
\begin{equation}
    \mathcal{E}_\text{rel}=\frac{\norm{\matr{F}-\tilde{\matr{F}}}_2}{\norm{\matr{F}}_2}\,,
\end{equation}
where $\matr{F}$ the reference data is given in \cref{eq:snapshotmatrix} and the reconstructed data $\tilde{\matr{F}}$ is either $\matr{F}_r$ in case of the POD or the FCNN predictions with $r$ latent variables for every $(x,t,c)$ in the data set.

\subsection{Singular value decay of reference data}
\begin{figure}[htbp!]
\centering
	\begin{subfigure}{.45\textwidth}
        \centering
		\begin{tikzpicture}

\begin{groupplot}[group style={group size=2 by 1, horizontal sep=1cm, vertical sep=2cm}]
\nextgroupplot[
log basis y={10},
tick align=outside,
tick pos=left,
x grid style={white!69.0196078431373!black},
xlabel={\(r\)},
xmin=-0.95, xmax=41.95,
xtick style={color=black},
y grid style={white!69.0196078431373!black},
ylabel={\(\sigma\)},
ymin=5.00425450746683e-16, ymax=233.601752047754,
ymode=log,
ytick style={color=black},
width=.6\textwidth,
height=.7\textwidth,
ytick={1e1,1e-3,1e-11,1e-15},
width=.55\textwidth,
height=.7\textwidth,
y label style={yshift=-2.5em},
grid=both
]
\addplot [semithick, red, mark=o, mark size=2, mark options={solid}]
table {%
1 36.7564464709921
2 6.45572768708393
3 2.70096946740729
4 0.706165973805752
5 0.280354602766687
6 0.102537716843524
7 0.0440340620127454
8 0.0218638471526527
9 0.00685281169963933
10 0.00206424289417759
11 0.00100028680918087
12 0.0002164070851145
13 0.000192016045990026
14 2.76177875569651e-05
15 1.65500681978358e-05
16 2.40203186862594e-06
17 1.13376480492873e-06
18 1.54303137841961e-07
19 6.12006889847331e-08
20 7.054966207969e-09
21 2.62311702138279e-09
22 2.48217436010154e-10
23 8.81706226450051e-11
24 6.88254148471367e-12
25 2.27942669067321e-12
26 1.42331417671071e-13
27 4.36913792303503e-14
28 1.63220898105638e-14
29 3.18040162440537e-15
30 3.18040162440537e-15
31 3.18040162440537e-15
32 3.18040162440537e-15
33 3.18040162440537e-15
34 3.18040162440537e-15
35 3.18040162440537e-15
36 3.18040162440537e-15
37 3.18040162440537e-15
38 3.18040162440537e-15
39 3.18040162440537e-15
40 3.18040162440537e-15
};

\nextgroupplot[
tick align=outside,
tick pos=left,
x grid style={white!69.0196078431373!black},
xlabel={\(r\)},
xmin=-0.95, xmax=41.95,
xtick style={color=black},
y grid style={white!69.0196078431373!black},
ylabel={cusum-e},
ymin=0.769785900167041, ymax=1.01096257618252,
ytick style={color=black},
ytick={1,.8},
width=.55\textwidth,
height=.7\textwidth,
y label style={yshift=-2em},
grid=both
]
\addplot [semithick, red, mark=o, mark size=2, mark options={solid}]
table {%
1 0.780748476349562
2 0.917875430648483
3 0.975247075592702
4 0.990246839199716
5 0.996201887904449
6 0.998379904929493
7 0.999315238187322
8 0.999779651018009
9 0.999925212485639
10 0.999969059338346
11 0.999990306561001
12 0.999994903292138
13 0.99999898193003
14 0.99999956856306
15 0.999999920105218
16 0.99999997112709
17 0.999999995209537
18 0.99999999848711
19 0.999999999787082
20 0.999999999936937
21 0.999999999992655
22 0.999999999997927
23 0.9999999999998
24 0.999999999999946
25 0.999999999999995
26 0.999999999999998
27 0.999999999999999
28 0.999999999999999
29 0.999999999999999
30 0.999999999999999
31 0.999999999999999
32 0.999999999999999
33 0.999999999999999
34 0.999999999999999
35 0.999999999999999
36 0.999999999999999
37 0.999999999999999
38 0.999999999999999
39 0.999999999999999
40 0.999999999999999
};
\addplot [thick, , mark=x,black, mark size=4, mark options={solid}]
table{%
4 0
4 0.990246839199716
4 1.3
};
\end{groupplot}

\end{tikzpicture}
		\subcaption{Hydrodynamic regime.}
		\label{Fig:CumSum_Rare}
	\end{subfigure}\hfill
	\begin{subfigure}{.45\textwidth}
        \centering
		\begin{tikzpicture}

\begin{groupplot}[group style={group size=2 by 1, horizontal sep=1cm, vertical sep=2cm}]
\nextgroupplot[
log basis y={10},
tick align=outside,
tick pos=left,
x grid style={white!69.0196078431373!black},
xlabel={\(r\)},
xmin=-0.95, xmax=41.95,,
xtick style={color=black},
y grid style={white!69.0196078431373!black},
ylabel={\(\sigma\)},
ymin=2.86160392849359e-16, ymax=240.32740800328,
ymode=log,
ytick style={color=black},
ytick={1e1,1e-3,1e-11,1e-15},
width=.55\textwidth,
height=.7\textwidth,
y label style={yshift=-2.5em},
grid=both
]
\addplot [semithick, red, mark=o, mark size=2, mark options={solid}]
table {%
1 36.8185958349281
2 5.78483852846218
3 2.9488881352441
4 1.08115123432794
5 0.4715894924307
6 0.27551553286601
7 0.155493855631619
8 0.0601331453526982
9 0.05155511017701
10 0.0132542951500055
11 0.0118122790965581
12 0.00208495452053553
13 0.00184461993287337
14 0.000261109297076443
15 0.000174118703867616
16 2.59262125849959e-05
17 1.30752219185821e-05
18 1.92140998809572e-06
19 9.1685066176197e-07
20 1.08651788755093e-07
21 5.26354986735524e-08
22 4.52124036969183e-09
23 2.44729256168519e-09
24 1.43747068296607e-10
25 9.28350136149776e-11
26 3.39879764456285e-12
27 2.74182349854538e-12
28 6.4789443879917e-14
29 6.12293316992491e-14
30 1.6313694307166e-14
31 2.92181471455653e-15
32 2.92181471455653e-15
33 2.92181471455653e-15
34 2.92181471455653e-15
35 2.92181471455653e-15
36 2.92181471455653e-15
37 2.92181471455653e-15
38 2.92181471455653e-15
39 2.92181471455653e-15
40 1.8678655154319e-15
};

\nextgroupplot[
tick align=outside,
tick pos=left,
x grid style={white!69.0196078431373!black},
xlabel={\(r\)},
xmin=-0.95, xmax=41.95,
xtick style={color=black},
y grid style={white!69.0196078431373!black},
ylabel={cusum-e},
ymin=0.760859233580134, ymax=1.0113876555438,
ytick style={color=black},
ytick={1,.8},
width=.55\textwidth,
height=.7\textwidth,
y label style={yshift=-2em},
grid=both
]
\addplot [semithick, red, mark=o, mark size=2, mark options={solid}]
table {%
1 0.772246889123937
2 0.893580238655189
3 0.95543131247198
4 0.978107779579132
5 0.987999072557982
6 0.993777837563443
7 0.997039223381299
8 0.998300478281949
9 0.999381814290284
10 0.999659814794242
11 0.999907569918492
12 0.999951300528145
13 0.99998999027099
14 0.999995466874262
15 0.999999118904556
16 0.999999662690558
17 0.999999936935114
18 0.99999997723548
19 0.999999996465846
20 0.999999998744749
21 0.999999999848745
22 0.999999999943575
23 0.999999999994906
24 0.999999999997921
25 0.999999999999868
26 0.999999999999939
27 0.999999999999997
28 0.999999999999998
29 0.999999999999999
30 1
31 1
32 1
33 1
34 1
35 1
36 1
37 1
38 1
39 1
40 1
};
\addplot [thick, , mark=x,black, mark size=4, mark options={solid}]
table{%
6 0
6 0.993777837563443
6 1.3
};
\end{groupplot}
\end{tikzpicture}
		\subcaption{Rarefied regime.}
		\label{Fig:CumSum_Hydro}
	\end{subfigure}
	\caption{Singular value decay \(\sigma\) and cumulative energy increase for the number of singular variables \(k\) in the hydrodynamic regime (a) and in the rarefied regime (b). A black cross marker corresponds to over 99\% cumulative energy. 
    }
	\label{Fig:CUSUM-e}
\end{figure}
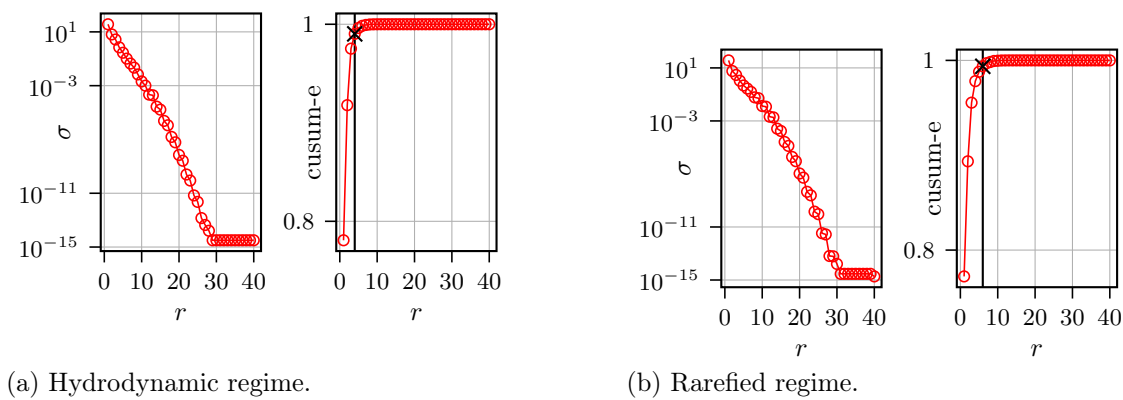

As a first step, we perform a POD with the hydrodynamic data and with the rarefied data. The obtained singular values \(\sigma\), as well as the cumulative energy (cusum-e) defined in \cref{e:ePOD}, are shown in \cref{Fig:CUSUM-e}. As expected, more modes are necessary in the rarefied regime compared to the hydrodynamic regime. With a total of \(\intrPOD=4\) intrinsic variables, a cumulative energy of over \(99\%\) can be achieved for the hydrodynamic regime. 
The cumulative energy of the singular values of the rarefied regime only reaches above \(99\%\) with \(\intrPOD=6\) singular values. 

For the POD we define the \textit{intrinsic dimension} $\intrPOD$ as the smallest truncation rank $r$ of the reduced system, at which the cumulative energy $E_\text{cum}$ defined in \cref{e:ePOD} reaches $99\%$. Although this choice is arbitrary it is a common practice in classical MOR to truncate \cref{eq:POD-approx}, whenever $99\%$ of the cumulative energy is reached.

In \cref{Fig:CUSUM-e}, we further see that the rate at which the singular values drop is approximately exponential in both regimes, which has been also observed by \cite{bernard2018reduced}. Consequently, a rapid decay of the Kolmogorov N-width is indicated. 
Note that the singular value decay is similar for both domains but not exactly the same, thus leading to an expected increase in the number of intrinsic variables in the rarefied regime necessary to achieve similar \(L_2\)-errors.

It is important to note that the parameter $\intrPOD$ is not expected to precisely match the actual intrinsic dimension $\intrDim$ of the solution manifold. The intrinsic dimension represents the minimum number of variables required to accurately describe the system's exact solution manifold. This discrepancy arises because the solution manifold is fundamentally nonlinear, making it challenging to adequately capture with a parsimonious linear approximation.

For the FCNN, the intrinsic dimension is defined as the smallest number of intrinsic variables that minimizes the error. In well-trained models, the FCNN's intrinsic dimension should ideally align with $\intrDim$.

From a fluid mechanics perspective, the hydrodynamic regime theoretically requires only $\intrDim=3$ intrinsic variables. This is because near-equilibrium flows in this regime can be effectively characterized by three macroscopic quantities: density $\rho$, macroscopic velocity $u$, and total energy $E$, as outlined in \cref{Eq:Conservation1}-\cref{Eq:Conservation3}, see also \cite{bernard2018reduced,Koellermeier2017a}.

Conversely, the rarefied regime demands a larger intrinsic dimension, denoted as $\intrDim$. This is due to the need for more than only the equilibrium Maxwellian distribution function to describe the microscopic velocities adequately. Therefore, we initially set $\intrDim=3$ intrinsic variables ($\idhy$) for the FCNN in the hydrodynamic case and choose $\intrDim=5$ intrinsic variables ($\idrare$) for the rarefied regime. This choice aligns with extended fluid dynamic models as described in \cite{Koellermeier2017a,Torrilhon2016}.

We note that each FCNN with different latent space dimension $r$ needs to be trained separately. This is different from the POD, where the decomposition is only performed once. Thereafter the approximation quality is given by the truncation rank $r$.

\subsection{Variation of the number of intrinsic variables}
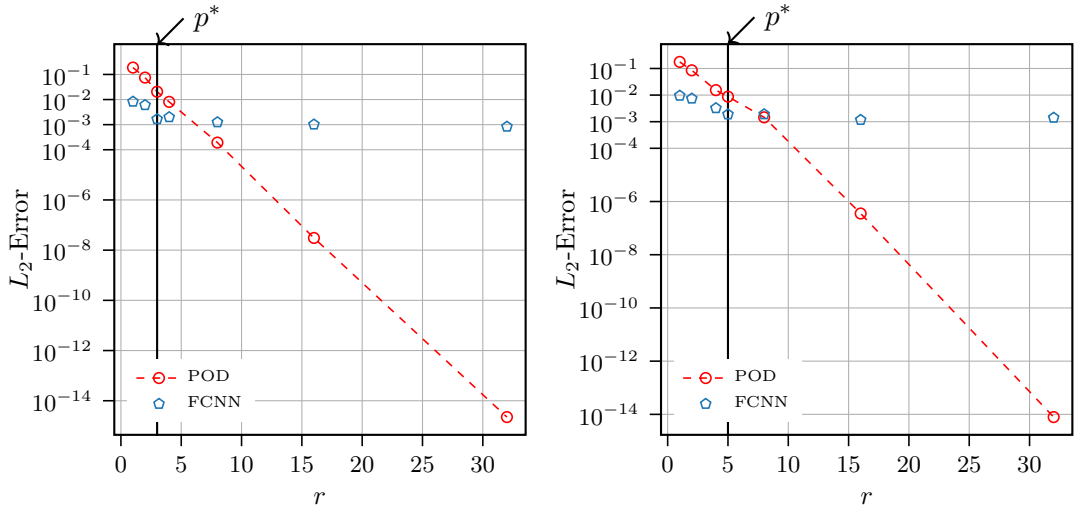
\begin{figure}[htbp!]
	\begin{tikzpicture}
\definecolor{color0}{rgb}{0.12156862745098,0.466666666666667,0.705882352941177}

\begin{groupplot}[group style={group size=2 by 1,horizontal sep=1.8cm}]
\nextgroupplot[
legend cell align={left},
legend style={draw=none,at={(0.03,0.03)}, anchor=south west},
log basis y={10},
tick align=outside,
tick pos=left,
x grid style={white!69.0196078431373!black},
xmajorgrids,
xmin=-0.55, xmax=33.55,
xminorgrids,
xtick style={color=black},
y grid style={white!69.0196078431373!black},
ymajorgrids,
ymin=4.38349387313967e-16, ymax=1.61134858880557,
yminorgrids,
ymode=log,
ytick style={color=black},
ytick={1e-1,1e-2,1e-3,1e-4,1e-6,1e-8,1e-10,1e-12,1e-14},
xlabel={{\(r\)}},
ylabel={\(L_2\)-Error},
width=0.47\textwidth,
height =.45\textwidth,
clip=false,
y label style={yshift=-.7em},
max space between ticks=20
]
\addplot [semithick, red, mark=o, mark size=2, mark options={solid}, dashed]
table {%
1 0.188112310801957
2 0.0750338979596223
3 0.020528730333796635
4 0.00808627149114823
8 0.000193252431896578
16 3.06183124046159e-08
32 2.23530701528268e-15
};
\addlegendentry{POD}
\addplot [semithick, color0, mark=pentagon, mark size=2, mark options={solid}, only marks]
table {%
1 0.00824632961302996
2 0.0060168607160449
3 0.0016505243
4 0.00198765122331679
8 0.00124555476941168
16 0.00101344427093863
32 0.000832061574328691
};
\addlegendentry{FCNN}
\draw[thick](3,40e-17)--(3,1.5);
\draw [thick,<-] (axis cs:3,1.5)-- +(10pt,10pt) node[right] {\(\intrDim\)};
\nextgroupplot[
legend cell align={left},
legend style={draw=none, at={(0.03,0.03)}, anchor=south west},
log basis y={10},
tick align=outside,
tick pos=left,
x grid style={white!69.0196078431373!black},
xmajorgrids,
xmin=-0.55, xmax=33.55,
xminorgrids,
xtick style={color=black},
y grid style={white!69.0196078431373!black},
ymajorgrids,
ymin=1.70008814466799e-15, ymax=0.821373329691319,
yminorgrids,
ymode=log,
ytick style={color=black},
ytick={1e0,1e-1,1e-2,1e-3,1e-4,1e-6,1e-8,1e-10,1e-12,1e-14},
xlabel={{\(r\)}},
ylabel={\(L_2\)-Error},
width=0.47\textwidth,
height =.45\textwidth,
clip=false,
y label style={yshift=-.7em},
max space between ticks=20
]
\addplot [semithick, red, mark=o, mark size=2, mark options={solid}, dashed]
table {%
1 0.176637499442346
2 0.0853532495802733
4 0.015335605212791
5 0.008731715326052242
8 0.00145958547754045
16 3.54159334428613e-07
32 7.90549608414532e-15
};
\addlegendentry{POD}
\addplot [semithick, color0, mark=pentagon, mark size=2, mark options={solid}, only marks]
table {%
1 0.00949728023260832
2 0.00740677583962679
4 0.0032303836196661
5 0.0018827654
8 0.00190045870840549
16 0.00116818200331181
32 0.00140941923018545
};
\addlegendentry{FCNN}
\draw[thick](5,16.5e-16)--(5,0.85);
\draw [thick,<-] (axis cs:5,0.85)-- +(10pt,10pt) node[right] {\(\intrDim\)};
\end{groupplot}

\end{tikzpicture}
	\caption{The $L_2$-Error over the variation of the latent space dimension/truncation rank \(r\) using FCNN/POD for the hydrodynamic regime (left) and the rarefied regime (right).}
	\label{Fig:IntVar}
\end{figure}

The variation of the number of intrinsic variables \(r\) in \cref{Fig:IntVar} sheds light on the performance of the autoencoder with different bottleneck layer sizes. In the case of the POD $r$ is the truncation rank of the decomposition \cref{eq:POD-approx} and the latent space dimension in case of the FCNN. To this end, $r$ is varied for both the POD and the FCNN over $r \in \{1,2,3,4,8,16,32\}$ for the hydrodynamic case and over $r \in \{1,2,4,5,8,16,32\}$ for the rarefied case. We note that the loss of information when applying POD goes exponentially to zero with increasing $r$, which is not surprising when consulting the \textit{Eckard-Young Theorem}  \cite{EcY1936}.
Note that the FCNN is retrained for each different $r$. By changing $r$, i.e. widening the bottleneck layer, a gain or loss of capacity occurs that can be connected to stability during training.

Both for the hydrodynamic and the rarefied regime, POD initially yields a larger error than the FCNN for small number of intrinsic variables $r$. Not surprisingly, the POD accuracy increases with the number of singular values taken into account until the error reaches machine precision. The FCNN error decreases as well and then reaches a plateau, with a typical remaining error due to the network architecture and training. For the previously identified values $\intrDim=3$ in the hydrodynamic case and $\intrDim=5$ in the rarefied case, the FCNN results in a more accurate approximation than the POD.

We note that when testing the FCNN against POD and fixing \(r\) the FCNN is limited by the estimation error of the training and performs under its abilities. However, POD uses five to six times more parameters than the FCNN while the deterministic character enables POD to achieve any possible accuracy, which was not observed with the neural network.

In the following, we consider the intrinsic variables with constant values $\intrDim=3$ in the hydrodynamic case and $\intrDim=5$ in the rarefied case.

This leads to the number of trainable parameters of the POD and the FCNN shown in \cref{Tab: Parameters}. We can see that the FCNN achieves a relatively small error with a small number of parameters in comparison with the POD for this choice of the number of intrinsic variables. 

\begin{table}[htp]
	\centering
	\caption{Amount of parameters used to reconstruct \(f\), the number of intrinsic variables \(p\) and the corresponding $L_2$-Error for POD and FCNN, both for hydrodynamic (\(\hy\)) and rarefied (\(\rare\)) regimes.}
	\begin{tabular*}{\textwidth}{ @{\extracolsep{\fill}} c c c c c c c @{} }
		\toprule
		Algorithm & \multicolumn{2}{c}{Parameters \(\frepar\)} & \multicolumn{2}{c}{Int. variables \(p\)}& \multicolumn{2}{c}{$L_2$-error} \\ [.5ex]
		 & \(\hy\)&\(\rare\)&\(\hy\)&\(\rare\)&\(\hy\)&\(\rare\)\\
		\hline
		POD     & 15129 & 25225 & 3 & 5 & 0.0205 & 0.0087 \\
		FCNN 	& 2683 & 3725 & 3 & 5 & 0.0008 & 0.0009 \\
		\bottomrule
	\end{tabular*} \label{Tab: Parameters}
\end{table}

Next, a qualitative analysis with the actual reconstructions is presented. From computations of the \(L_2\)-error over time \(t\), which are not shown due to conciseness, it became clear that the time step that contributes most to the error is the last time step in case of POD, while the FCNN distributes the error more evenly over all time steps.

\subsection{Reconstruction quality}
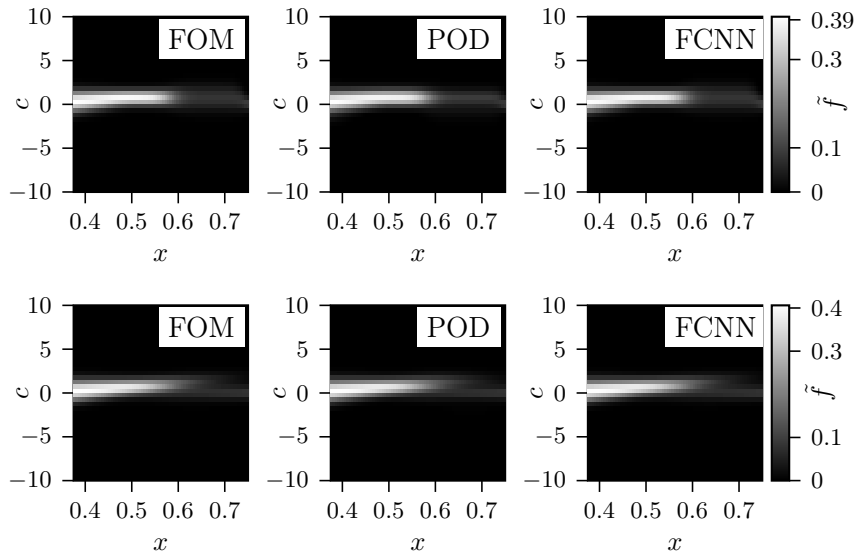
\begin{figure}[htp!]
    \centering
	\begin{tikzpicture}

\begin{groupplot}[
group style={group size=3 by 2,
	horizontal sep= 1.1cm,
	vertical sep = 1.5cm},
tick align=outside,
tick pos=left,
x grid style={white!69.0196078431373!black},
xmin=0.375, xmax=0.75,
xtick style={color=black},
y grid style={white!69.0196078431373!black},
ymin=-10, ymax=10,
ytick style={color=black},
height=.26\textwidth,
width=.26\textwidth,
xlabel={\(x\)},
ylabel={\(c\)},
y label style={yshift=-1.4em}
]
\nextgroupplot[
]
\addplot graphics [
includegraphics cmd=\pgfimage,
xmin=0.375, xmax=0.75,
ymin=-10, ymax=10
] {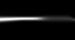};
\node[fill=white] at (axis cs:0.65,7.5) {FOM};
\nextgroupplot[
]
\addplot graphics [
includegraphics cmd=\pgfimage,
xmin=0.375, xmax=0.75,
ymin=-10, ymax=10
] {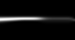};
\node[fill=white] at (axis cs:0.65,7.5) {POD};
\nextgroupplot[
colorbar,
colorbar style={
	ylabel={$\tilde{f}$},
	ytick={0,0.1,.3,.39},
	yticklabels={0,0.1,0.3,0.39},
	y label style={yshift=1.3cm},
	ticklabel style={font=\footnotesize},
	tick align=outside,
	tick pos=right,
	width=0.1*\pgfkeysvalueof{/pgfplots/parent axis width},
	xshift=-0.2cm
},
colormap/blackwhite,
point meta max=0.397007430832041,
point meta min=8.50982895819395e-73,
]
\addplot graphics [
includegraphics cmd=\pgfimage,
xmin=0.375, xmax=0.75,
ymin=-10, ymax=10
] {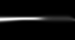};
\node[fill=white] at (axis cs:0.65,7.5) {FCNN};
\nextgroupplot[
]
\addplot graphics [
includegraphics cmd=\pgfimage,
xmin=0.375, xmax=0.75,
ymin=-10, ymax=10
] {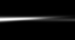};
\node[fill=white] at (axis cs:0.65,7.5) {FOM};
\nextgroupplot[
]
\addplot graphics [
includegraphics cmd=\pgfimage,
xmin=0.375, xmax=0.75,
ymin=-10, ymax=10
] {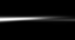};
\node[fill=white] at (axis cs:0.65,7.5) {POD};
\nextgroupplot[
colorbar,
colorbar style={
	ylabel={$\tilde{f}$},
	ytick={0,0.1,.3,.4},
	yticklabels={0,0.1,0.3,0.4},
	y label style={yshift=1.3cm},
	ticklabel style={font=\footnotesize},
	tick align=outside,
	tick pos=right,
	width=0.1*\pgfkeysvalueof{/pgfplots/parent axis width},
	xshift=-0.2cm
},
colormap/blackwhite,
point meta max=0.406101604565777,
point meta min=1.26406789996295e-34
]
\addplot graphics [
includegraphics cmd=\pgfimage,
xmin=0.375, xmax=0.75,
ymin=-10, ymax=10
] {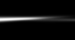};
\node[fill=white] at (axis cs:0.65,7.5) {FCNN};
\end{groupplot}

\end{tikzpicture}
	\caption{Comparison of the FOM solutions \(f\) (left column) with reconstructions \(\tilde{f}\) obtained from POD (middle column) and FCNN (right column) at end time \(t=0.12s\) for \(x\in [0.375,0.75]\) in the hydrodynamic regime (top row) and in the rarefied regime (bottom row).}
	\label{Fig: ErrWorst}
\end{figure}

The reconstructed solutions compared to the full order model (FOM) at the final time step $t=0.12s$ are  given in \cref{Fig: ErrWorst} . Because of the small overall errors indicated in \cref{Tab: Parameters} both the POD and the FCNN reproduce the FOM solution without any visible differences at first sight.

\begin{figure}[htb]
	\input{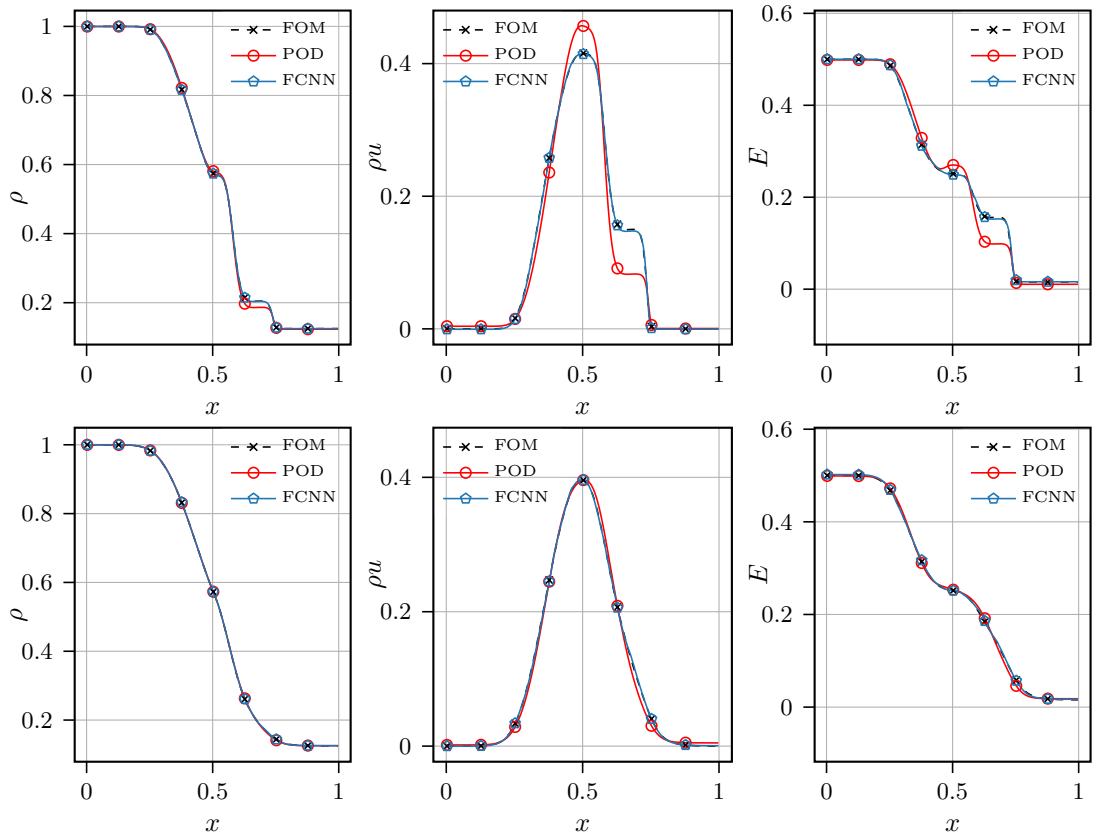}
	\caption{Comparison of FOM macroscopic quantities \(\rho\), \(\rho u\) and \(E\) with POD and the FCNN for the hydrodynamic regime (top row) and the rarefied regime (bottom row) at time \(t=0.12s\).}
	\label{Fig:ErrMacro}
\end{figure}

As shown in \cref{Fig:ErrMacro}, the more subtle information loss from the model reduction can unfold in actual differences in the macroscopic quantities \(\rho\), \(\rho u\) and \(E\). 
Overall, \cref{Fig:ErrMacro} shows that the errors are larger for the hydrodynamic regime (top row), most notably for the momentum and energy of the POD model close to the contact discontinuity. However, the position of the shock is well approximated. In contrast, the FCNN model yields a very good agreement in the hydrodynamic case.
For the rarefied regime, both models approximate the FOM solution very well. The lack of sharp shock structures in the full model and the increased intrinsic dimension $\intrDim=5$ combined seem to notably influence the accuracy.

\subsection{Conservation properties}
The physical consistency of the reduced \(\tilde{f}\), in terms of conservation of mass, momentum, and energy, is a critical criterion for its validity. Hence, conservation properties are analyzed in the following. We note that conservation of mass, momentum, and energy is not directly built-in by means of a specifically tailored loss function. Even though we can expect to recover some conservation properties as they are implicitly built into the numerical reference data. The investigation of different loss functions to improve upon this is left for future work. 

We investigate the conservation properties by means of the derivative of the cell-averaged conserved quantities mass, momentum, and total energy, defined exemplary as
\begin{equation}
	\frac{\mathrm{d}}{\mathrm{d}t}\int \rho(x,t)\, \mathrm{d}x\Delta t  =\overline{\dot{\rho}}\mathrm{,}
\end{equation}
for the mass. Note that a derivative $\overline{\dot{\rho}}= 0$ denotes conservation of mass, for example. We expect the conservation of mass and total energy, while the momentum increases with a constant rate, due to the boundary conditions of the test case, featuring a larger pressure on the left-hand side of the domain.

\begin{figure}[htb]
	\input{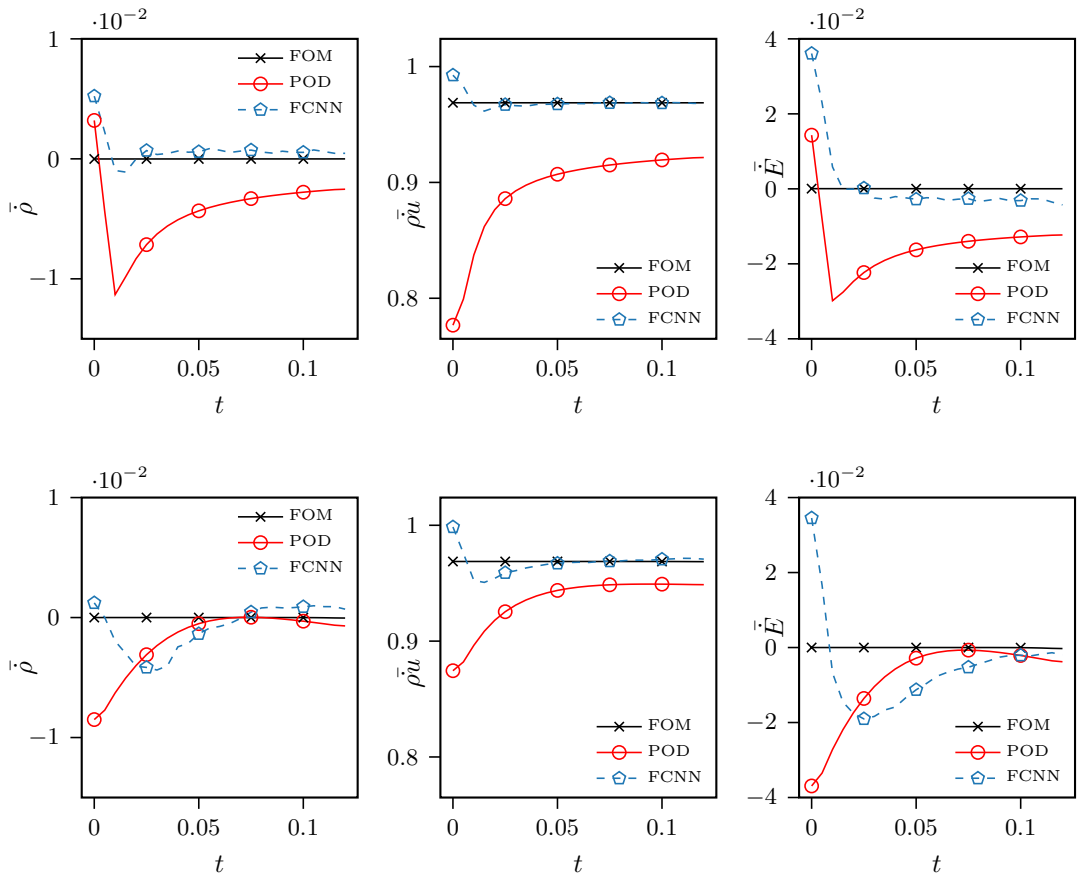}
	\caption{Comparison of the conservative properties of reconstructions obtained from POD and the FCNNs with the conservative properties of the FOM solution using the mass, momentum, and energy derivative. A value of $0$ indicates conservation. FCNN approximately conserves mass and energy, while the momentum increases with the correct rate of the test case.}
	\label{Fig:Conservation}
\end{figure}

\Cref{Fig:Conservation} shows the evolution of the derivatives of mass, momentum, and total energy as a function of time for the hydrodynamic regime (top row) and the rarefied regime (bottom row).
Indeed, \cref{Fig:Conservation} indicates that conservation of mass and energy are achieved with reasonable accuracy for the FCNN, while the error is larger for the POD reconstruction for both regimes. Also the increase in momentum of the full order model (FOM) is accurately described by the FCNN with a larger error for the POD method for both regimes. Overall, the errors are slightly smaller for the rarefied case compared to the hydrodynamic case, which might be due to the higher capacity of the neural network and more modes for the POD using five intrinsic variables in the rarefied regime.

\subsection{Physical interpretability}
An important question in the context of model reduction with neural networks is the interpretability of the results, because of the usual black-box nature of neural networks \cite{fan2021interpretability}. 
Especially when benchmarking neural networks for model order reduction with POD, evaluating the interpretability of the intrinsic variables is important, since POD by construction achieves a so-called physically interpretable decomposition of the input data \cite{Kutz2019} as outlined previously. 

Following the assumption that the hydrodynamic case can be fully described in terms of three macroscopic quantities and that the rarefied case is reasonably describable in a similar way with an extended set of five variables, we test the intrinsic variables \(\idhy\) and \(\idrare\) for similarities and investigate if they match any macroscopic quantity. Two related macroscopic quantities, namely the temperature \(T\) and macroscopic velocity \(u\), are added to the three macroscopic variables $\rho$, $\rho u$, and $E$. In \cref{Fig: Macro_hy} and \cref{Fig: Macro_rare}, these are plotted first over the whole domain of \(x\) and \(t\) and for the end time \(t=0.12s\) for both regimes. Similarly, we plot both the FCNN's and the POD's first 3 intrinsic variables \(\idhy\) of the hydrodynamic case and 5 intrinsic variables \(\idrare\) of the rarefied test case
\begin{align*}
    \idhy &= [h_0(x,t),h_1(x,t),h_2(x,t)],\\
    \idrare &= [r_0(x,t),r_1(x,t),r_2(x,t),r_3(x,t),r_4(x,t)],
\end{align*}
depicted in \cref{Fig: Code_hy} and \cref{Fig: Code_rare} respectively.

\begin{figure}[htp!]
	\centering
	\input{fom_mac_hy.tex}
	\caption{Macroscopic quantities of the hydrodynamic case obtained by the FOM. Density \(\rho\), momentum \(\rho u\), total energy \(E\), temperature \(T\), and velocity \(u\) over time \(t\) and space \(x\) in the top row and at \(t=0.12\) in the bottom row.}
	\label{Fig: Macro_hy}
	
    \input{code_hydro.tex}
	\caption{Intrinsic variables \(h_0(x,t)\), \(h_1(x,t)\) and \(h_2(x,t)\) of hydrodynamic case obtained by the FCNN. Top row depicts the whole \((x,t)\) domain, bottom row is for \(t=0.12\).}
	\label{Fig: Code_hy}

    \setlength\figureheight{0.35\linewidth}%
    \setlength\figurewidth{0.5\linewidth}%
	\input{hydro_POD1.tex}
	\caption{First three intrinsic variables \(h_0(x,t)\), \(h_1(x,t)\) and \(h_2(x,t)\) of hydrodynamic case obtained by the POD. Top row depicts the whole \((x,t)\) domain, bottom row is for \(t=0.12\).}
	\label{Fig:POD_hy}
\end{figure}

\begin{figure}[hbp!]
\centering
	\input{fom_mac_rare.tex}
		\caption{Macroscopic quantities of the rarefied case obtained by the FOM. Density \(\rho\), momentum \(\rho u\), total energy \(E\), temperature \(T\), and velocity \(u\) over time \(t\) and space \(x\) in the top row and at \(t=0.12\) in bottom row.}
	\label{Fig: Macro_rare}

	\input{code_rare.tex}
	\caption{Intrinsic variables \(r_0(x,t)\), \(r_1(x,t)\), \(r_2(x,t)\), \(r_3(x,t)\) and \(r_4(x,t)\) of rarefied case obtained by the FCNN. Top row depicts the whole \((x,t)\) domain, bottom row is for \(t=0.12\).}
	\label{Fig: Code_rare}

    \setlength\figureheight{0.35\linewidth}%
    \setlength\figurewidth{0.5\linewidth}%
	\input{POD_rare.tex}
	\caption{First five intrinsic variables \(r_0(x,t)\), \(r_1(x,t)\), \(r_2(x,t)\), \(r_3(x,t)\) and \(r_4(x,t)\) of rarefied case obtained by the POD. Top row depicts the whole \((x,t)\) domain, bottom row is for \(t=0.12\).}
	\label{Fig: POD_rare}

\end{figure}

Strikingly, most intrinsic variables of the FCNN in \cref{Fig: Code_hy} and \cref {Fig: Code_rare} and the POD in \cref{Fig:POD_hy} and \cref {Fig: POD_rare} appear to be a combination of the five intrinsic variables shown in \cref{Fig: Macro_hy} and \cref{Fig: Macro_rare}, respectively. 
In particular consider FCNN in the hydrodynamic case by comparing \cref{Fig: Macro_hy} and \cref{Fig: Code_hy}. The rarefaction wave, shock wave, and contact discontinuity, which can be identified in \(h_0\) reflect a combination of those found in the density \(\rho\) and the total energy \(E\). Furthermore, \(h_1\) seems to model the negative momentum \(\rho u\), with different boundary values. The temperature \(T\) appears linked to \(h_2\), where the same fluctuation appears. 
Similar results hold true for the POD variables in \cref{Fig:POD_hy}, where especially the first two variables closely resemble the density $\rho$ and the momentum $\rho u$. Interestingly, this does not relate to very good conservation properties of the POD in comparison with FCNN as shown in the previous section and \cref{Fig:Conservation}. 
Considering the FCNN in the rarefied case, we compare \cref{Fig: Macro_rare} and \cref{Fig: Code_rare}. Here \(r_3\) clearly reflects the shape of the density \(\rho\). Moreover, the peak of the velocity \(u\) can be observed in \(r_0\). For the other intrinsic variables of \(r_1\), \(r_2\) and \(r_4\) a clear discernability of macroscopic quantities is difficult to observe and might require linear or nonlinear combinations. Additionally, it is possible that those intrinsic variables resemble non-equilibrium variables not present among the macroscopic variables. Considering the POD results in \cref{Fig: POD_rare}, we again see a very good agreement of the first intrinsic variables with density and momentum.

For more physical insight into the relation between macroscopic variables and intrinsic variables, the Pearson correlation between all variable combinations is computed in \cite{Goodfellow-et-al-2016} for the FCNN. Note that we expect similar results for the POD based on the previous results, but do not present them here for conciseness. The Pearson correlation coefficient $r_{X,Y} = r_{Y,X}$ is a measure of linear correlation between two sets of data, here represented by a macroscopic variable $X \in \{\rho, \rho u, E, T, u\}$ and an intrinsic variable $Y \in \{r_0, r_1, r_2,\}$ for the hydrodynamic case and $Y \in \{r_0, r_1, r_2, r_3, r_4\}$ for the rarefied case. It is commonly defined as 
\begin{equation}
    r_{X,Y} = \frac{\sum \left(x_i - \Bar{x} \right)\left(y_i - \Bar{y} \right)}{\sum \sqrt{\left(x_i - \Bar{x} \right)}\sqrt{\sum \left(y_i - \Bar{y} \right)}}.
\end{equation}
Note that $r_{X,Y}\in [-1,1]$, with $r_{X,Y} = 0$ meaning that there is no correlation between both data sets, $r_{X,Y} = 1$ indicating a perfect correlation, and $r_{X,Y} = -1$ indicating a perfect anti-correlation.

\begin{figure}
\centering
	\input{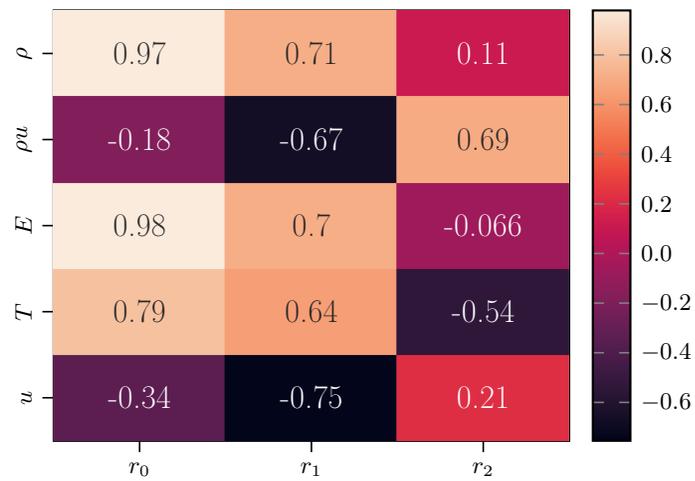}
	\caption{Pearson correlation between macroscopic quantities and intrinsic variables for the hydrodynamic case.}
    \label{fig:pearson_hy}
\end{figure}

The Pearson correlation coefficients for the hydrodynamic test case are presented in \cref{fig:pearson_hy}. As predicted by the previous analysis, there appears to be an almost perfect correlation of the first intrinsic variable $r_0$ with the density $\rho$. This means that the FCNN succeeds at identifying the density $\rho$ precisely as an internal variable. Note that $r_0$ also correlates almost perfectly with the energy $E$, as the energy depends linearly on $\rho$. Additionally, there is a relatively strong linear correlation between $r_2$ and both the momentum $\rho u$ as well as the temperature $T$. The intrinsic variable $r_2$ on the other hand seems to be correlated with all variables. 

\begin{figure}
    \centering
	\input{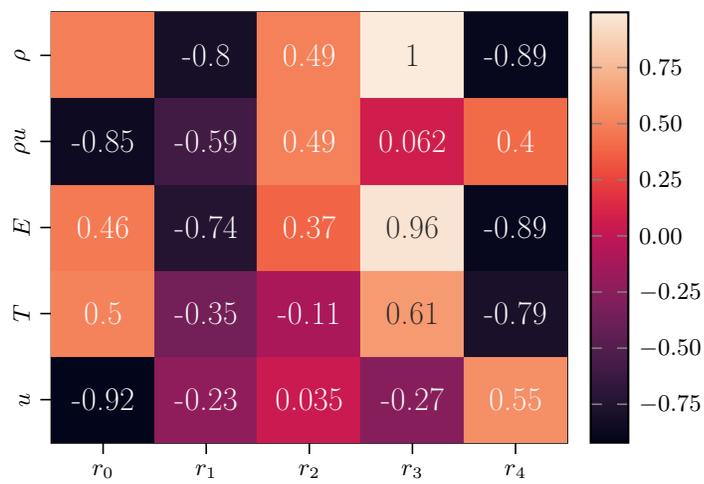}
	\caption{Pearson correlation between macroscopic quantities and intrinsic variables for the rarefied case.}
    \label{fig:pearson_rare}
\end{figure}

The Pearson correlation coefficients for the rarefied test case are presented in \cref{fig:pearson_rare}. In agreement with the previous results, there is no clear correlation of most of the intrinsic variables. An exception is $r_0$, which is anti-correlated with $u$ and $r_3$, which correlates almost perfectly with the energy $E$. Note that only linear correlations are tested here. For further analysis on the disentanglement of the intrinsic variables, it might be more suitable to consider nonlinear correlations beyond the Pearson correlation coefficients. One option that could be explored in future work would be to train the functional relation between the intrinsic variables and macroscopic variables, for example by means of symbolic regression \cite{Cranmer2023}.

\section{Conclusion}
\label{sec:conclusion}

This paper marks the first comparison of velocity space model reduction techniques for rarefied flows using proper orthogonal decomposition (POD) and a fully connected neural network (FCNN).

As physically expected, the rarefied regime needs more modes than the hydrodynamic regime. Choosing three and five intrinsic variables for the hydrodynamic and rarefied case, respectively, leads to less than one percent error.
The FCNN is initially more accurate but has a remaining error even for larger number of intrinsic variables, while POD achieves subsequently higher accuracy  with increasing the number of parameters.
The resulting errors of the macroscopic variables are small, especially in the smoother rarefied case.

Even though not strictly enforced, the FCNN approximately exhibits the correct conservation of mass, momentum, and energy, while POD has a slightly larger error.

The correlation of intrinsic variables and macroscopic variables is investigated by means of the evolution of the reconstructed values and the pairwise correlation for the FCNN. The density is directly included in the latent space while the relation with other macroscopic variables seems to be more complex.

In addition to optimizing the neural network's performance, our research endeavors to enhance its predictive capabilities by incorporating fundamental physical properties such as conservation and interpretability. This objective includes the exploration of different loss functions and the regularization techniques.
Furthermore, future work can utilize the model reduction approach in real-world scenarios, including simulations and parameter predictions.
Lastly, the presented concepts could be applied in related fields, such as shallow flows \cite{Koellermeier2020c} or fusion plasmas \cite{KrahYinBergmannNaveSchneider2023}.

\section*{Author Contribution Statement (CRediT)}
All authors contributed to this publication.
The authors contributions differ in the following points:

\vspace{5pt}
{\small
\noindent
\begin{tabular}{@{}lp{10.8cm}}
\textbf{Julian Koellermeier:} & initial idea, methodology, setup of test cases, reference data, writing\\
\textbf{Philipp Krah:} & initial idea, methodology, implementation, visualization, writing\\
\textbf{Julius Reiss:} & initial idea, supervision \\
\textbf{Zachary Schellin:} & implementation, training, visualization\\
\end{tabular}
}

\section*{Conflict of Interest}
The authors declare that they have no conflict of interest.

\section*{Code and Data Availability}
All data and scripts to reproduce the results can be obtained from the authors on reasonable request.

\section*{Acknowledgement}
The authors would like to acknowledge the financial support of the CogniGron research center and the Ubbo Emmius Funds (University of Groningen). 
Centre de Calcul Intensif d’Aix-Marseille is acknowledged for granting access to its high performance computing resources.
P. Krah acknowledges partial funding from the Agence Nationale de la Recherche (ANR), project CM2E, grant ANR-20-CE46-0010-01.

\bibliographystyle{abbrv}
\bibliography{references}

\appendix
\renewcommand{\L}{L_}
\section*{Appendix: Hyperparameters for the Fully Connected Autoencoder}
\label{appendix}
In this section, we describe the tests that have been conducted to optimize our hyperparameters. The hyperparameters include: the number of layers, i.e., depth; the number of nodes per hidden layer, i.e., width, batch size, and non-linear activation functions; the number of epochs for training; and the learning rate. Experiments are evaluated through: the validation error which estimates the model's ability to generalize; the training error which estimates the optimization to training data; %
and the $\L2$-error, which gives an estimate of how well the model performs on the whole dataset and hence is applied as a comparative metric against POD.

To start with a working model, an estimate over the initial hyperparameters is done, which are summarized in \Cref{Tab:First Guess}. These include a mini-batch size of 16, the width of the bottleneck layer is 3 and 5  for the hydrodynamic case \(\hy\) and the rarefied case \(\rare\), respectively, and a learning rate of 0.0001. LeakyReLU is applied as an activation function for the output, input, and any hidden layer besides the output of the bottleneck layer, and is referred to as activation hidden. The hyperbolic tangent is applied as an activation function for the output of the last hidden layer in the encoder which outputs the code, referred to as the activation code. Moreover, 2000 initial number of epochs are used. This might appear exaggerated but is justified by the small amount of input data and the small size of the network which yields fast training.
\begin{table}[htp!]
	\centering
	\caption{The initial selection of batch size, bottleneck size, number of epochs, learning rate, and applied activation functions.}
	\begin{tabular}{ c c c c c }
		\toprule
		Mini-batch size   & Intr. dim. &   Epochs &Learn. rate & Activ. hidden/code \\ [.5ex]   
		\midrule
		16 		&	3/5 &     2000&	    0.0001 & LeakyReLU/Tanh\\
		\bottomrule
	\end{tabular} \label{Tab:First Guess}
\end{table}

\begin{figure}[htbp!]
  \includegraphics[width=0.9\textwidth]{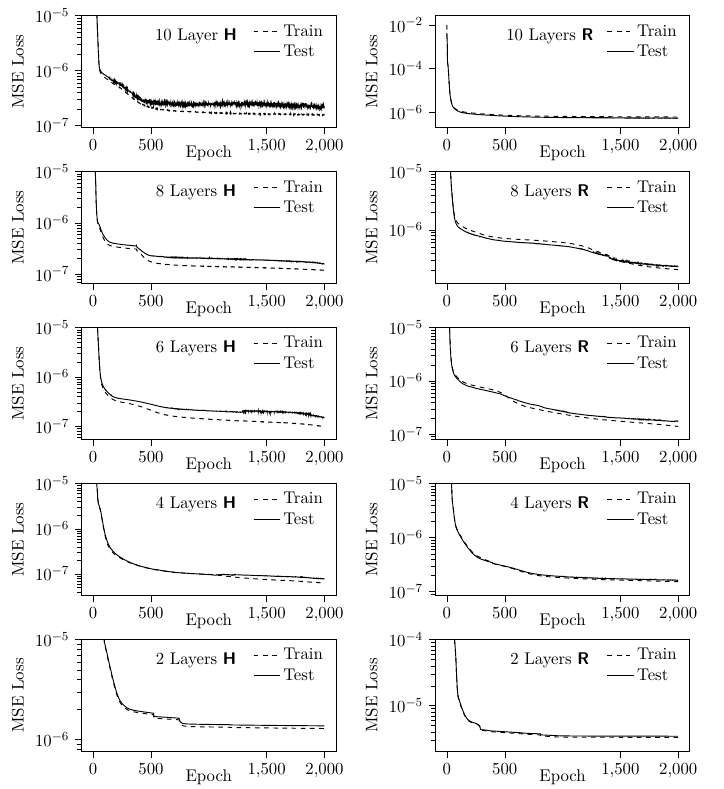}
		\caption{Five experiments over the different depths with the hydrodynamic case $\hy$ left and the rarefied case $\rare$ right. The number of layers used for every experiment is given. Training and validation loss are shown over 2000 epochs.}
		\label{Fig:Depth}
	\end{figure}

Five designs for finding an optimal number of layers, i.e., depth, are explored. These are as follows:     
\begin{enumerate}
	\item 10 layers with layer widths: 40, 40, 20, 10, 5, 3/5, 5, 10, 20, 40, 40.
	\item 8 layers with layer widths: 40, 40, 20 , 10, 3/5, 10, 20, 40, 40.
	\item 6 layers with layer widths: 40, 40, 20 , 3/5, 20, 40, 40.
	\item 4 layers with layer widths: 40, 40, 3/5, 40, 40.
	\item 2 layers with layer widths: 40, 3/5, 40.
	\end{enumerate}
The model's depth is determined in a primary step because it determines the model's representational capacity and therefore can initiate over- and underfitting at an early stage in the hyperparameter search. The results of the experimentation are shown in \cref{Fig:Depth} and \cref{Tab:Depth} for both rarefaction levels.

\begin{table}[htp]
	\centering
	\caption{Results for the variation of the depth. Given are minimum values of training and validation error as well as the \(\L2\) error. The minima were reached around the last 50 epochs of the training. The \(\L2\) error is evaluated with the model at the last epoch.}
	\begin{tabular*}{15cm}{ @{\extracolsep{\fill}} c c c c c c c @{} }
		\toprule
		Depth & \multicolumn{2}{c}{Minimum training error} & \multicolumn{2}{c}{Minimum validation error} & \multicolumn{2}{c}{\(\L2\) error }\\ [.5ex]
		 & \(\hy\)&\(\rare\)&\(\hy\)&\(\rare\)&\(\hy\)&\(\rare\)\\
		\midrule
		10& \num{1.53e-7} & \num{5.96e-7} & \num{2.22e-7} & \num{5.19e-7} & 0.0048 & 0.0091\\
		8 & \num{1.17e-7 }& \num{2.05e-7} & \num{1.58e-7} & \num{2.32e-7} & 0.0041 & 0.0054\\
		6 & \num{9.76e-8} & \num{1.40e-7} & \num{1.49e-7} & \num{1.72e-7} & 0.0038 & 0.0045\\
		4 & \num{6.29e-8} & \num{1.52e-7} & \num{7.74e-8} & \num{1.61e-7} & 0.0031 & 0.0048 \\
		2 & \num{1.29e-6} & \num{3.29e-6} & \num{1.37e-6} & \num{3.42e-6} & 0.0136 & 0.0217\\ 
  \bottomrule
	\end{tabular*}\label{Tab:Depth}
\end{table}\noindent

For the hydrodynamic case \(\hy\), the lowest validation error of \num{7.74e-8} and an \(\L2\) error of 0.0031 is reached with 4 layers and constitutes the best-performing design. Additionally, as seen in \cref{Fig:Depth}(left), a design that exceeds 4 layers results in slight overfitting from the 500th epoch. Less than 4 layers do not reach the validation error and \(\L2\) error of the other designs, yielding the conclusion, that the capacity is too low. Overfitting occurs with 4 layers only after the 1000th epoch and is of smaller magnitude compared to the other three models that show overfitting.

For rarefied case \(\rare\), the lowest validation error of \num{1.61e-7} is also reached with 4 layers. On the other hand, the lowest \(\L2\) error of 0.0031 and the lowest training error of \num{1.40e-7} are reached with 6 layers. Contrary to the previously discussed hydrodynamic case, the training error and \(\L2\) errors are of lower magnitude for 6 layers, except for the validation error. Looking at \cref{Fig:Depth}(right), it is observed that the model with 6 layers starts to overfit after the 1500 epochs, yielding a decreasing training error and a stagnating validation error. Hence the model improved in the optimization task which additionally improves the \(\L2\) error. Its generalization ability, measured by the validation error, did not improve and is larger than the validation error reached with 4 layers. This concludes a model with 4 layers constitutes the best-performing.

Qualitatively, the overall training for both rarefaction levels is very stable. Training and validation errors do not diverge excessively and converge early in training. Separation of training and validation error occurs prominently for the hydrodynamic solution.

\begin{figure}[htbp!]
		\includegraphics[width=0.9\textwidth]{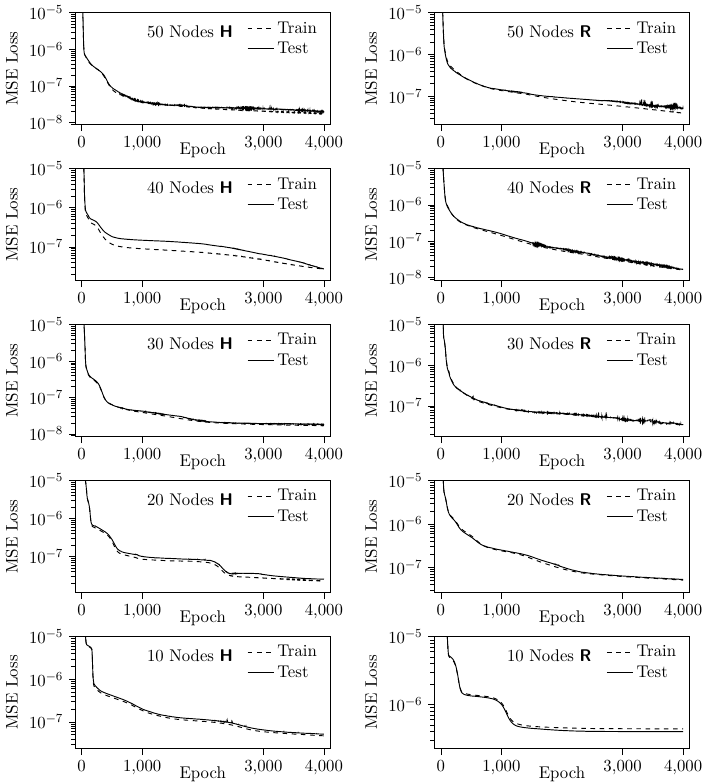}		
		\caption{Five experiments over different widths with the hydrodynamic case $\hy$ left and the rarefied case $\rare$ right. The number of nodes used for every experiment is given. Training and validation loss are shown over 4000 epochs.}
		\label{Fig:Width}
	\end{figure}

The width of the two remaining hidden layers is examined in the following. For both the hydrodynamic and the rarefied regime five experiments are conducted, lowering the hidden units of the hidden layers from fifty to ten. Note that the decoder is chosen to be structurally a reflection of the encoder. Therefore only one parameter is changed. Results for the hydrodynamic case \(\hy\) and the rarefied case \(\rare\) are shown in \cref{Tab:Width}. Note that the contribution of over-and underfitting is negligible and therefore the training error is omitted. A model with 30 hidden units in encoder and decoder performs best with the hydrodynamic case \(\hy\) and reaches a validation error of \(\num{1.77e-08}\). The corresponding \(\L2\) error is equal to \( \num{1.5e-3}\) with a shrinkage factor of 0.015. Overall, the loss of each experiment with the hydrodynamic case \(\hy\) is quite similar and ranges from \(\num{1.77e-8}\) to \(\num{5.11e-8}\). The \(\L2\) error behaves similarly and is even equal for 50 and 30 layers. A model with 40 hidden units performs best for the rarefied case \(\rare\). The corresponding validation error is \(\num{1.65e-8}\) with \(\L2=\num{1.4e-3}\), which is smaller than for the hydrodynamic case \(\hy\). The shrinkage factor here is 0.125. In all experiments, a model with 10 hidden nodes performs worst. Training and validation errors over 4000 epochs for both experiments can be seen in \cref{Fig:Width}.

\begin{table}[htpb!]
	\centering
	\caption{Results for the variation of the width. Given are the minimum value of the validation error as well as the \(\L2\) error. The minima were reached around the last 50 epochs of the training, the \(\L2\) error is evaluated with the model at the last epoch.}
	\begin{tabular*}{15cm}{ @{\extracolsep{\fill}} c c c c c c c @{} }
		\toprule
		Hidden units & \multicolumn{2}{c}{Validation error} & \multicolumn{2}{c}{$\L2$} & \multicolumn{2}{c}{Shrinkage factor}\\ [.5ex]
		& \(\hy\)&\(\rare\)&\(\hy\)&\(\rare\)&\(\hy\)&\(\rare\)\\
		\hline
		50 & \num{1.91e-8}  & \num{5.05e-8} & \num{0.0015}  & \num{0.0025} & 0.06  & 0.01\\
		40 & \num{2.65e-08} & \num{1.65e-8} & \num{0.0018}  & \num{0.0014} & 0.075 & 0.125\\
		30 & \num{1.77e-08} & \num{3.40e-8} & \num{0.0015}  & \num{0.0021} & 0.015 & 0.0167\\
		20 & \num{2.50e-08} & \num{5.25e-8} & \num{0.0017}  & \num{0.0027} & 0.1   & 0.25 \\
		10 & \num{5.11e-08} & \num{3.97e-7} & \num{0.0025}  & \num{0.0077} & 0.3   & 0.5\\\bottomrule
	\end{tabular*}\label{Tab:Width}
\end{table}

\begin{figure}[htbp!]
    \includegraphics[width=0.9\textwidth]{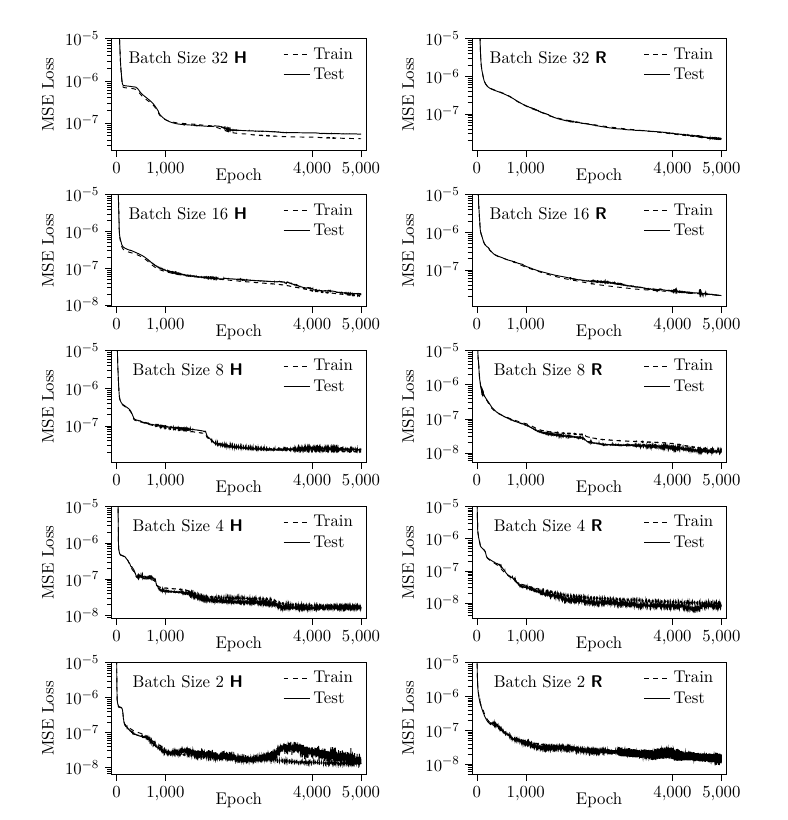}
		\caption{Five experiments over different batch sizes with the hydrodynamic case $\hy$ left and the rarefied case $\rare$ right. The batch size used for every experiment is given. Training and validation loss are shown over 5000 epochs.}
		\label{Fig:batch}
\end{figure}

 Next, the mini-batch size is analyzed. Results are displayed in \cref{Tab:Batch}. Experiments are conducted with mini-batch sizes of 2, 4, 8, 16, 32. The smallest batch size of 2 yields the lowest validation error of \(\num{1.15e-8}\) with corresponding \(\L2 = 0.0012\) at epoch 4956 for the hydrodynamic case \(\hy\). The lowest validation error with \(\num{6.30e-9}\) is achieved for the rarefied case \(\rare\) at epoch 4534 with a batch size of 4. The corresponding \(\L2\) error equals  \(0.001\). Small batch sizes have a regularizing effect on the training and therefore are beneficial to generalization. At the same time, the lower the batch size is, the more unstable is the training as seen in \cref{Fig:batch}. The oscillations that begin with batch sizes of 8 and lower, which make the training unstable, can be cured with a lower learning rate as soon as training starts to tremble. Additionally, small batch sizes drastically increase training time, thus a batch size as low as 2 is not used for the experiments. In conclusion, a batch size of 4 is chosen. Furthermore, a reduction of the learning rate from \(\num{1e-4}\) to \(\num{1e-5}\) is applied after the 3000th epoch.
 
\begin{table}[htp!]
	\centering
	\caption{Results for the variation of the batch sizes. Given are the minimum value of validation error as well as the corresponding epoch. Additionally, the \(\L2\) error is given but evaluated with the model at the last epoch.}
	\begin{tabular*}{15cm}{ @{\extracolsep{\fill}} c c c c c c c @{} }
		\toprule
		Batch Size & \multicolumn{2}{c}{Validation error} & \multicolumn{2}{c}{$\L2$} &\multicolumn{2}{c}{Epoch}\\ [.5ex]
		& \(\hy\)&\(\rare\)&\(\hy\)&\(\rare\)&\(\hy\)&\(\rare\)\\
		\hline
		32& \num{5.40e-8} & \num{2.17e-8} & \num{0.0024}  & \num{0.0017}&4998&4992\\
		16& \num{1.95e-8} & \num{2.06e-8} & \num{0.0015}  & \num{0.0016}&4999&5000\\
		8 & \num{2.25e-8} & \num{1.03e-8} & \num{0.0017}  & \num{0.0012}&4965&4961\\
		4 & \num{1.52e-8} & \num{6.30e-9} & \num{0.0013}  & \num{0.0010}&3956&4534\\
		2 & \num{1.15e-8} & \num{9.18e-9} & \num{0.0012}  & \num{0.0013}&4956&4872\\\bottomrule
	\end{tabular*}\label{Tab:Batch}
\end{table}

Eight experiments with different activation functions, ReLU, ELU, Tanh, SiLU, and LeakyReLU, are performed. The experiment designs and results are given in \cref{Tab:activations} for hidden and code activations. With the hydrodynamic case \(\hy\), combining ELU and ELU for hidden and code activation, respectively, yields the best result for the validation error with \num{4.44e-9} and a corresponding \(\L2\) error of 0.0008. These values are achieved at the last epoch. For the rarefied case \(\rare\), a combination of ReLU and ReLU for hidden and code activation, respectively, produces a validation error of \num{7.18e-9} and a corresponding \(\L2\) error of 0.0009. Both are reached close to the last epoch. Note that all models reach their lowest loss at or very close to the last epoch. The reason is the stable training after the 3000th epoch, where the learning rate is lowered to \num{1e-5} as seen in \cref{Fig:Activations}. This measure shows in all experiments an immediate success for learning. Both validation and training errors fall at the 3001st epoch and only decrease slightly thereafter. This behavior clearly shows that the updates to the free parameters \(\frepar\) were too big, which prohibitively slowed down or even prevented the learning process. Small updates to \(\frepar\) made all models quickly reach a minimum.

\begin{table}[htp!]
	\centering
	\caption{Results for the variation of the activation functions for the hidden-/code layers. Given are the minimum value of validation error as well as the corresponding epoch and the \(\L2\) error. The \(\L2\) error is evaluated with the models saved when the minimum validation error was achieved during training.}
	\begin{tabular*}{15cm}{ @{\extracolsep{\fill}} c c c c c c c @{} }
		\toprule
		Activ. hidden/code & \multicolumn{2}{c}{Validation error} & \multicolumn{2}{c}{$\L2$} &\multicolumn{2}{c}{Epoch}\\ [.5ex]
		& \(\hy\)&\(\rare\)&\(\hy\)&\(\rare\)&\(\hy\)&\(\rare\)\\
		\midrule
		ReLU/ReLU 	       & \num{9.79e-9} & \num{7.18e-9} & \num{0.0010}  & \num{0.0009}&5000 &4998\\
		ELU/ELU            & \num{4.44e-9} & \num{1.11e-8} & \num{0.0008}  & \num{0.0012}&5000 &5000\\
		Tanh/Tanh 	       & \num{7.83e-9} & \num{2.58e-8} & \num{0.0011}  & \num{0.0018}&5000 &5000\\
		SiLU/SiLU 	       & \num{7.69e-9} & \num{1.37e-8} & \num{0.0011}  & \num{0.0013}&5000 &5000\\
		LeakyReLU/LeakyReLU& \num{1.86e-8} & \num{9.39e-9} & \num{0.0015}  & \num{0.0010}&5000 &4997\\
		ELU/Tanh           & \num{5.49e-9} & \num{1.87e-8} & \num{0.0008}  & \num{0.0014}&5000 &5000\\
		LeakyReLU/Tanh     & \num{1.00e-8} & \num{1.42e-8} & \num{0.0010}  & \num{0.0012}&4997 &4992\\
		ELU/SiLU           & \num{8.11e-9} & \num{1.93e-8} & \num{0.0011}  & \num{0.0015}&5000 &5000\\\bottomrule
	\end{tabular*}\label{Tab:activations}
\end{table} 
The final hyperparameters for both input data are summarized below in \cref{tab:autoencoder-params}. From the initial models to the final models, the decrease in the validation error gained \(\approx \num{1.5e-7}\) for hydrodynamic case \(\hy\) and \(\approx \num{7.2e-8}\) for rarefied case \(\rare\) which amount to 93\% of the initial values for both models.


\begin{figure}[htp]
	\includegraphics[width=0.9\textwidth]{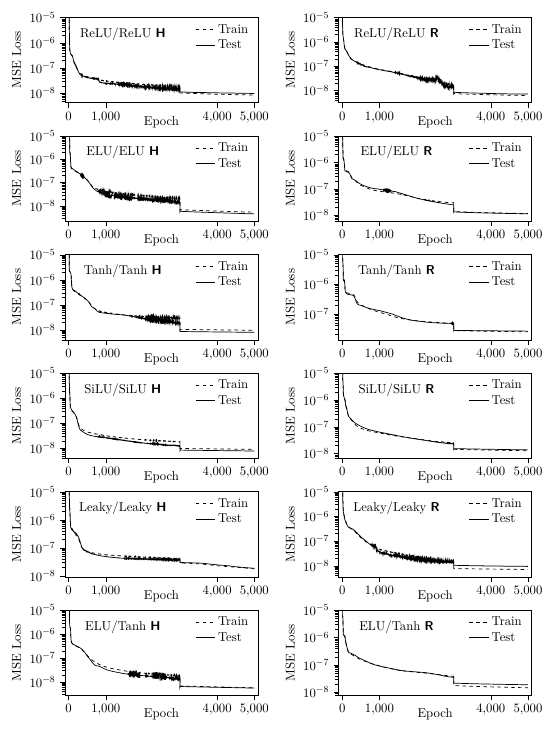}	
    \caption{Eight experiments with different combinations of activation functions for the hydrodynamic case $\hy$ left and the rarefied case $\rare$ right. Shown are training- and validation error over 5000 epochs.}
\end{figure}\label{Fig:Activations}

\end{document}